\documentclass[twocolumn,showpacs,amsmath,amssymb,prd,nofootinbib]{revtex4}
\usepackage{epsfig}
\usepackage[section]{placeins}
\def\sss{\scriptscriptstyle}
\def\_#1{_{\sss #1}}

\def\beq{\begin{equation}}
\def\eeqno#1{\label{#1}\end{equation}}

\def\az{a_0}
\def\A{\mathcal{A}}

\def\azg{\A_0}
\def\baz{\bar a_0}

\def\rar{\rightarrow}
\def\s{\sigma}
\def\r{\rho}

\def\mss{{\rm m~s^{-2}}}
\def\cmss{{\rm cm~s^{-2}}}

\def\S{\Sigma}

\def\f{\phi}
\def\grad{\vec\nabla}

\def\gf{\grad\phi}
\def\gfN{\grad \f_N}

\def\vr{ \textbf{r}}

\def\vv{\textbf{v}}

\def\vg{\textbf{g}}

\def\M{\mathcal{M}}

\def\b{\beta}
\def\c{\gamma}
\def\d{\delta}

\def\gN{g\_N}
\def\vgN{\vg\_N}

\def\Sbz{\S^{\sss 0}\_B}
\def\Spz{\S^{\sss 0}_p}

\def\SM{\Sigma\_M}

\def\rM{r\_ M}
\def\lM{\ell\_ M}
\def\lU{\ell\_U}

\def\dsr{\ell_{{\small \Lambda}}}

\newcommand{\LCDM}{$\Lambda$CDM~}

\begin{document}

\title{MOND vs. dark matter in light of historical parallels}
\author{Mordehai Milgrom }
\affiliation{Department of Particle Physics and Astrophysics, Weizmann Institute, Rehovot Israel 7610001}

\begin{abstract}
MOND is a paradigm that contends to account for the mass discrepancies in the Universe without invoking `dark' components, such as `dark matter' and `dark energy'.
It does so by supplanting Newtonian dynamics and General Relativity, departing from them at very low accelerations.
Having in mind readers who are historians and philosophers of science, as well as physicists and astronomers, I describe in this review the main aspects of MOND -- its statement, its basic tenets, its main predictions, and the tests of these predictions -- contrasting it with the dark-matter paradigm. I then discuss possible wider ramifications of MOND, for example the potential significance of the MOND constant, $\az$, with possible implications for the roots of MOND in cosmology.
Along the way I point to parallels with several historical instances of nascent paradigms. In particular, with the emergence of the Copernican world picture, that of quantum physics, and that of relativity, as regards their initial advent, their development, their schematic structure, and their ramifications. For example, the interplay between theories and their corollary laws, and the centrality of a new constant with converging values as deduced from seemingly unrelated manifestations of these laws. I demonstrate how MOND has already unearthed a number of unsuspected laws of galactic dynamics (to which, indeed, $\az$ is central) predicting them a priori, and leading to their subsequent verification. I parallel the struggle of the new with the old paradigms, and the appearance of hybrid paradigms at such times of struggle. I also try to identify in the history of those established paradigms a stage that can be likened to that of MOND today.
\end{abstract}


\maketitle

\section{Introduction}
Standard dynamics -- Newtonian and general-relativistic -- fail to account for the observed motions in galactic systems of all types,\footnote{Such as galaxies, dwarf satellites of galaxies, galaxy groups, and galaxy clusters.} and in the Universe at large, if gravity is generated only by the matter we have observed in these systems:
The measured accelerations in such systems are typically several to tens of times higher than what the visible matter can produce with standard gravity.
\par
The general acceptance of the reality of these discrepancies has lead to the emergence and entrenchment of the dark-matter (DM) paradigm.
According to this mainstream paradigm, familiar forms of matter -- such as in stars, dead or alive, gas, etc. collectively called `baryons' -- generate only a small part of gravity in galactic systems. The dynamical discrepancies are attributed to the presence of large amounts of `dark matter' whose extra gravity bridges the discrepancies.
\par
In the general budget of the Universe at large, DM makes up about twenty seven percent of the energy (or `mass'), compared with about five percent in baryons (and a modicum of other familiar species, such as electromagnetic radiation and neutrinos). The balance of about sixty eight percent is `dark energy'.
\par
While the ratio of DM to baryons in the universe is about 5:1, it can be much higher in galactic systems, where it can reach even 50-100 to 1. For recent reviews of the DM paradigm and its history see, e.g., Peebles, 2017 and Bertone \& Hooper, 2018.
\par
We know that the DM, if it exists, cannot be made of baryons, because there are not enough of these by a large margin (see e.g., Peebles, 2017). In fact, DM cannot be made of any constituents that are part of the standard model of particle physics -- all such candidates that were considered in the past have been ruled out. In other words,  {\it DM cannot be made of anything that is known to exist}.
\par
In contradistinction, MOND (Milgrom 1983, 1983a)\footnote{initially standing for `Modified Newtonian Dynamics', but now attaining a rather wider scope.} posits that the `dark matter' of the DM paradigm does not exist.\footnote{We know that some baryonic matter that is still `dark' is yet to be detected, because the observed baryons still do not tally with the amount of baryons we think exist.} Instead, MOND attributes the mass anomalies to unwarranted extrapolation of standard dynamics to the realm of the galaxies, where they have not been independently tested.
A relatively early review MOND can be found in Sanders \& McGaugh, 2002, and more recent ones in Famaey \& McGaugh, 2012, and in Milgrom, 2014 (the latter, an online review, is continually updated).
\par
\begin{figure}[!ht]
\includegraphics[width=1.0\columnwidth]{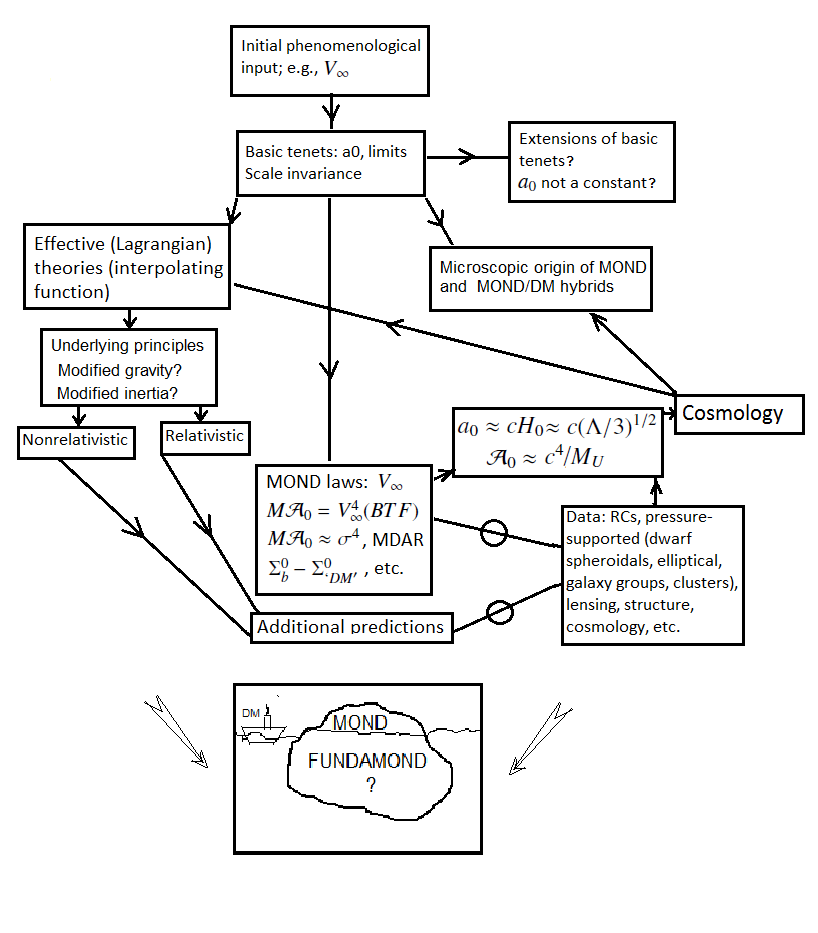}
\caption{A flowchart showing some aspects of the present status of activity, results, and desiderata within the MOND paradigm, to be unfolded in the present review.\label{flow}}
\end{figure}

In a way of a preview, I show in Fig. \ref{flow} a flowchart that tries to capture schematically the present state of the MOND paradigm, as described in the following sections.
\par
The present contribution is meant at familiarizing MOND to science historians and philosophers. But even veteran MOND practitioners will hopefully find novel insights in this more general view of MOND looked at from new angles.
\par
Partly to this end, I point extensively to
historical analogues that focus on, and limelight, certain aspects of MOND.
I see the main value in looking at such analogies in their giving us a guiding perspective on MOND, where we are still in the middle of the fray, and where it is not always easy to assess and attach significance to various findings: How seriously are we to take this or that possible conflict of observations with MOND, or the present inability of the paradigm to account for some phenomena? And, how do we weigh the successes of MOND against such difficulties? Have other, now-established, paradigms undergone similar questioning phases?
\par
The historical analogies I point to are practically all well-known, and for the most part I mention them without giving detailed references.
\par
I write from the viewpoint of a MOND advocate; so this article is not meant as a balanced presentation of the MOND-vs.-DM paradigm struggle. Reviews advocating the opposite stance abound.
\par
Earlier discussions of MOND vs. DM from the historical-philosophical perspective can be found in Sanders, 2010, 2016; Lahav \& Massimi, 2014; Merritt, 2016; Lazutkina, 2017; Massimi, 2018; and  Merritt, 2020.
\par
MOND is based on the following insight:
The accelerations in galactic systems are many orders of magnitude smaller than what we encounter on Earth or in the solar system. This makes it possible to conceive of a generalization -- MOND -- that departs from standard dynamics at very low accelerations, but preserves the known successes of these dynamics which all pertain to high accelerations.
\par
MOND thus introduces a new constant, $\az$, with the dimensions of acceleration. It then posits that standard dynamics are a good approximation at accelerations much above $\az$, but greatly amends them at low accelerations, comparable with, or smaller than, $\az$.
\par
MOND, if basically correct, would thus not abolish standard dynamics -- as the Copernican view abolished the Ptolemaic view. Rather, it would leave standard dynamics as a limit that works to high accuracy in some range of phenomena -- in the same way as  quantum theory and relativity relate to the 19th century classical dynamics.
\par
However, MOND, if established, would abolish DM -- a well-entrenched, mainstream paradigm that purports to account for the same phenomena as MOND, and to which many theorists and experimentalists cleave.
In this regard MOND's position is more akin to that of the Copernican paradigm in the first century after {\it `De revolutionibus'}. To my mind, it is mainly this fact that makes it an uphill struggle for MOND.
\par
Quantum physics, as a counterexample, has not had a competing paradigm to vie with in explaining `quantum phenomena'. Its struggle had been largely due to its perceived `weirdness' in various respects, which took years to be generally assimilated.
\par
Section \ref{brief} reviews some basic facts about the MOND paradigm, such as its basic axioms and some immediate consequences, comparing them with the general schemes of quantum theory and relativity.
Section  \ref{laws}  describes in more detail some of the major predictions of MOND, and how they stood the test of observations.
Section  \ref{DM} describes briefly what I consider some of the weak points of the DM paradigm.
I conclude with Sec. \ref{timeline}, which attempts to put MOND's present state in the context of established paradigms when they were at a similar state.

\section{MOND\label{brief}}
\subsection{ MOND -- basic tenets\label{tenets}}
MOND is based on the following basic tenets.
\par
1. A nonrelativistic, MOND-based theory of dynamics (gravity/inertia) should involve a new constant,
$\az$, with the dimensions of acceleration (beside $G$ and other standard-dynamics constants and parameters, such as masses).
We do not know whether MOND is relevant for non-gravitational dynamics; so I confine the discussion to gravity alone.
\par
2. A MOND theory should be well-approximated by standard dynamics when all acceleration-dimensioned characteristics of a system are much larger than $\az$ (in order to retain the successes of standard dynamics). Formally, this can be expressed as the requirement that any MOND expression containing $\az$ should tend to the corresponding standard one in the limit $\az\rar 0$. It follows, in particular, that $\az$ does not appear in the description of any high-acceleration phenomenon.
\par
3. The opposite, low-acceleration, or so-called deep-MOND limit (DML), is defined formally as the limit $\az\rar \infty$ (i.e., all characteristic accelerations are much smaller than $\az$). However, this limit has to be accompanied by taking, at the same time, the limit $G\rar 0$, such that
$\azg\equiv G\az$ is kept fixed (see Sec. \ref{cons} below for what is special in $\azg$).
\par
\begin{figure}[!ht]
\begin{center}
\includegraphics[width=0.9\columnwidth]{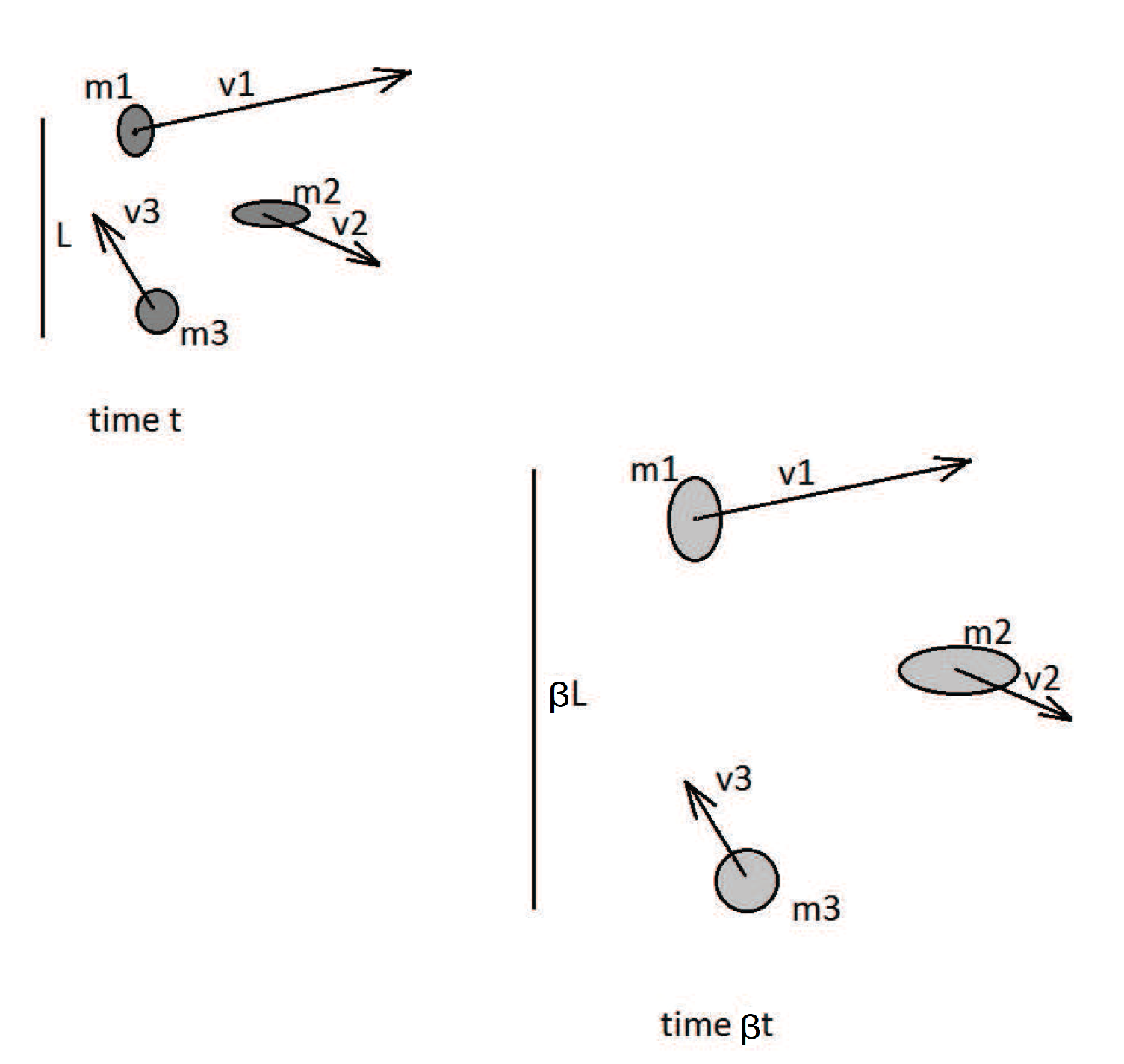}
\end{center}
\caption{The meaning of scale invariance: Two many-body systems, with their full time history, that are related to each other by space-time scaling by a factor $\b$.
All distances in the second system, at time $\b t$, are $\b$ times those in the first system at time $t$. All velocities at these two times are the same for each of the bodies comprising the system. Thus, the positions in the two systems are related by $\hat \vr_i(\b t)=\b\vr_i(t)$, or $\hat\vr_i(\hat t)=\b\vr_i(\hat t/\b)$, and $\hat\vv_i(\hat t)=\vv_i(\hat t/\b)$. \label{scaling}}
\end{figure}

The major new assumption of MOND, besides the introduction of $\az$, is that the DML is space-time scale invariant (Milgrom, 2009). This means that the DML equations are invariant under scaling of all times and all distances by the same factor $\b$; namely, under $(t,\vr)\rar \b(t,\vr)$ (see Fig. \ref{scaling}).
As is the case with symmetries of physical theories in general, this underlying symmetry implies that applying the symmetry operation to any solution of the theory yields another solution. A solution in the present context means the full time history of a gravitating system made of masses, such as a galaxy. Note, importantly, that velocities remain the same under space-time scaling.

\par
4. As an additional, small-print tenet we add, in deference to parsimony, the requirement that no other new dimensional constants appear in MOND. This means, in particular, that there are no very small or large dimensionless parameters involved.\footnote{If such a dimensionless parameter $\omega$ appear, then $\omega\az$, which would differ greatly from $\az$, would be a second acceleration constant.} One important result of this is that the transition from standard dynamics to the DML occurs not only around $\az$, but also within an acceleration range of order $\az$.

\subsubsection{Some consequences of the basic tenets\label{cons}}
Scale invariance (Milgrom , 2009), the symmetry principle that underlies the DML, has powerful consequences.
\par
Note first that
Newtonian dynamics, where the equations of motion in gravitating systems are of the schematic form  $a=MG/r^2$, is not scale invariant, since under scaling the right-hand side is multiplied by $\b^{-2}$, while the left-hand side is multiplied by $\b^{-1}$.
\par
The DML can be {\it schematically} described as `modified inertia' whereby Newton's 2nd law is replaced by
 $ma^2/\az=F$, but gravity remains intact: $F=mMG/r^2$ or as `modified gravity', where $ma=F$ still holds, but in gravity $F\propto m(MG\az)^{1/2}/r$. Both these schemes are scale invariant because combining the two we get in both cases $a= (M\azg)^{1/2}/r$.
\par
If all degrees of freedom in a system transform under scaling according to their $[m][\ell][t]$ dimensions,\footnote{Namely, a quantity of dimensions $[m]^a[\ell]^\c[t]^\d$ is multiplied by $\b^{\c+\d}$.} which can be assumed without loss of generality, then the only constants that can appear in a scale-invariant theory describing this system must have dimensions that leave them unchanged under the scaling of time and length by the same factor. (Note that when we apply a scaling transformation we only scale degrees of freedom, not the constants.) Under such scaling of the units, the value of $\az$ does change by a factor of $\b^{-1}$, and that of $G$ by a factor of $\b$. Thus, $\az$ and $G$ cannot appear in the DML, only the constant $\azg\equiv G\az$.
\par
One might invert the order of introducing the basic tenets of MOND, and of its two constants $\az$ and $\azg$. We can start by postulating that galactic dynamics in the region where large mass anomalies are found is governed by a scale-invariant theory. MOND is thus assumed to have two limiting dynamics: the $G$-controlled, standard dynamics, and the $\azg$-controlled, scale-invariant dynamics. Then, from $G$ and $\azg$ one constructs the constant that marks the boundary between the two limits, which gives rise to $\az=\azg/G$. See Milgrom, 2015a for the logic of following this route.
\subsection{Boundary constants and their roles -- analogies with quantum theory and relativity\label{analog}}
As regards the basic tenets, MOND is schematically similar to quantum physics and relativity in relation to the classical paradigms they replaced. As in MOND's first tenet, these last two also revolve around a dimensioned constant.
\par
In quantum theory, it was the Planck constant, $h$, having the dimensions of action (or angular momentum), and ushered in with the paradigm itself, as was $\az$ ($h$ was called `the quantum of action' by Planck ).
For relativity, it is $c$, the `speed of light' (which plays more fundamental roles than being the speed of propagation of light).
\par
In both cases, as in MOND's second tenet, it was required from the outset that the new paradigms tend to the classical theory in the previously explored regime. For quantum theory, this is anchored in Bohr's correspondence principle, which states that quantum expressions tend to the classical ones in the formal limit $\hbar\rar 0$ (I use $\hbar\equiv h/2\pi$ hereafter).
Also, general relativity (hereafter, GR) was constructed with the explicit requirement that it tends to Newtonian gravity for nonrelativistic phenomena -- formally when $c\rar\infty$ everywhere.
\par
It can be said then that $\hbar,~~c$, and $\az$ all play a role of `boundary constants', or delimiters of the applicability regime of the old paradigm, in whose equations they do not appear.\footnote{These `boundary constants' differ in this role from other constants appearing in physics, such as Boltzmann's $k$, and Newton's $G$, which do not mark boundaries, but are rather `conversion constants'. For example, $k$ is used to convert temperature to energy and would have not been needed had it been realized sooner that temperature stands for energy. Newton's $G$ converts gravitational masses to inertial ones, and has a meaning only if the ratio of the two is indeed universal.} Classical Newtonian dynamics is the common limiting paradigm of all three when $\hbar\rar 0,~~c\rar\infty,~~\az\rar 0$.
\par
Besides their role as `boundary constants, all three appear ubiquitously in the description of phenomena in the `new' regime beyond the boundary. As said above, in the case of MOND it is rather the constant $\azg=\az G$ that always appears in MOND's `new' regime, the DML.
\par
The ubiquitous appearance of $\az$ in the DML is further discussed in
Sec. \ref{lawsanalog}.

\section{MOND laws of galactic dynamics \label{laws}}
Newtonian dynamics is a general theory that enables us to calculate the evolution of any (nonrelativistic) gravitating system, given its initial state.
So should a MOND-based theory.
\par
We can extract from Newtonian dynamics a number of restricted but useful laws, such as Kepler's laws of planetary motions, the dependence of Kepler's constant on the mass of the central star,
the virial relation, etc.
\par
MOND too, even its basis tenets alone, predicts some concrete, restricted laws that focus on specific aspects of galactic dynamics.
As in any general theory, it is important and useful to extract such corollaries.
Such laws highlight simply formulated and memorable predictions that cut through the complexity of the full theory.
\par
Many characteristics of a galaxy -- such as its mass, size, composition, etc. -- depend strongly on its generically complicated and unknowable formation and evolution history. The MOND laws should however be obeyed independent of this
history.
I emphasize this (and see more in Sec. \ref{prediction} below) because in the context of the DM paradigm, the regularities encapsuled in these MOND laws are {\it none of them} laws of nature, as Kepler's laws are. Any regularities of this type would need to somehow result from the complicated and unknowable histories of individual galactic systems.
\par
Most of these MOND laws were listed in Milgrom (1983, 1983a); but a few were recognized later on. These are discussed in more detail in Milgrom (2014a).
\par
In Sec. \ref{examples}, I briefly list and comment on some of these laws, leaving for Sec. \ref{testlaws} the discussion of how they fare vis a vis the data. In the rest of this section, I only list some generalities regarding these laws, and also digress to make some comparisons with quantum theory and relativity.
\subsection{MOND laws -- generalities}
The MOND laws to be discussed below, follow, with a little leeway, from only the basic tenets of MOND (as detailed in Milgrom, 2014a).
\par
With few exceptions and caveats, these laws are independent as phenomenological laws -- e.g., if interpreted as effects of DM. By this I mean that one can construct imaginary models of galaxies with baryons and DM that will appear to satisfy any set of these laws
but not others.
\subsubsection{MOND mimics with dark matter\label{mimic}}
Even if it is MOND, and not the putative DM, that underlies the dynamical discrepancies, one may try to interpret some MOND results as the effects of DM. For example, if MOND predicts a certain gravitational field, $\vg(\vr)=-\gf$ for a given baryonic mass distribution, $\r\_B(\vr)$, while Newtonian gravity gives a field $\vgN=-\gfN$. Then one can invoke a total mass distribution $\r^*(\vr)$ that would give $\vg(\vr)$ with Newtonian dynamics, and interpret the excess $\r_p=\r^*-\r\_B$ as the required density of DM, and $\vg_p=\vg-\vgN$ as the acceleration produced by DM. Some of the MOND predictions may then be expressed as properties of such a phantom MOND-mimic DM.
\par
Not all of the MOND predictions can, however, be mimicked with DM, even given the very wide and extensive leeway afforded by this paradigm (see Sec. \ref{discriminants}).
\par
This leads to another general point about MOND laws:
if one interprets these laws as properties of phantom DM in galaxies, then some of these laws would correspond to properties of the phantom DM alone, some to those of the baryons alone, and some to relations between the two components. I shall exemplify this below, where I list some of the laws.

\subsection{examples of MOND laws\label{examples}}
Following are some of the salient MOND laws. Most were identified in Milgrom (1983, 1983a); for others I indicate the references below. (The order and numbering of these laws have no special significance.)

\par
(i) Asymptotic constancy of orbital velocity: The speed, $V(r)$, on a circular orbit of radius $r$, around an isolated central (baryonic) mass, $M$,  becomes independent of $r$ for large $r$: $V(r)\rar
V\_{\infty}$ as $r\rar\infty$. Here, `large enough' means a. far from the main body so it can be treated as a point mass, and b. we are already in the DML; namely, $r\gg \rM$, where $\rM$ is the `MOND radius' associated with the mass $M$:
\beq \rM\equiv \left(\frac{MG}{\az}\right)^{1/2}. \eeqno{mondradius}
The MOND radius for a mass $M$, which combines $MG$ with $\az$, is analogous to the Schwarzschild radius of a mass in the context of relativity, which combines $MG$ with $c$: $r\_S\equiv 2MG/c^2$. The former marks the radius of transition from the Newtonian regime (much below $\rM$) to the DML, as the latter marks the radius near which Newtonian dynamics gives way to GR.
\par
This law, which generalizes Kepler's 3rd law,
follows straightforwardly from the scale invariance of the DML: scaling the orbital radius and orbital time by the same factor does not change the orbital speed.
\par
If interpreted in terms of DM, it defines a property of the DM halo alone, since the rotation asymptote is dominated by the purported contribution of the DM to gravity.
\par
(ii) The extension of law (i) to light-bending predicts (again, from scale invariance of the DML) that the bending angle becomes asymptotically constant.
\par
(iii) The mass-asymptotic-speed relation (MASR):\footnote{Sometimes referred to as the baryonic Tully-Fisher relation (BTFR).} MOND predicts that the asymptotic speed and light-bending angle in laws (i-ii) depend only on the mass of the central body as\footnote{MOND predicts proportionality in this relation, and the normalization in the definition of $\az$ is conventionally defined so that equality holds. Sometimes the `best' value of $\az$ is determined from full rotation-curve analysis of galaxies, not from the MASR alone. But these give consistent values.}
\beq V\_{\infty}^4=M\azg,   \eeqno{masr}
and
\beq \theta=2 V^2\_{\infty}/c^2=2(M\azg)^{1/2}/c^2.  \eeqno{bending}
\par
In terms of DM, this would be a relation between a DM property ($V\_{\infty}$) and a baryon property ($M$).
\par
This law has no analogue in Kepler's laws themselves -- which were enunciated for only the solar system. It generalizes the dependence of Kepler's constant on the central mass, which emerged from Newtonian dynamics.
\par
(iv) In a self gravitating, isolated system in the DML, the characteristic intrinsic speed, $\sigma$, and the total mass, $M$, are related by a `virial relation':
\beq \sigma^4=qM\azg, \eeqno{msigma}
where $q$ is of order 1. In a large class of MOND theories, $q=4/9$ (Milgrom, 2014b). This contrasts with the analogue virial relation in Newtonian dynamics, which involves also the size of the system, $R$, and reads: $\sigma^2\sim MG/R$.

\par
The fact that system size does not appear in the DML virial relation is, again, a result of scale invariance.
\par
In terms of DM, this law is a relation between the DM (which determines $\sigma$ as it dominates over baryons for DML systems) and baryons ($M$).
\par
The DML laws (i)-(iv) are analogues of Newtonian laws.
But, some MOND laws have no analogues in Newtonian dynamics, since they pertain to the transition, around $\az$, from the DML to Newtonian dynamics.
In these laws appears not $\azg$, but $\az$ itself, or its proxy, the MOND surface density (or column density)
\beq \SM\equiv \frac{\az}{2\pi G}.   \eeqno{sigmond}
\par
(v) The mass anomaly appears (e.g., in rotation-curve analysis) always around the radius $R$ where  $V^2/R=\az$.
This transition occurs at $\rM$ if the mass is well within $\rM$, but not so if the mass extends beyond $\rM$.
There are galaxies in which $V^2/r\ll\az$ everywhere, for which MOND has predicted that the discrepancy should exist at all radii.
\par
(vi) The excess acceleration in MOND over what Newtonian dynamics dictate -- namely, the `phantom' acceleration, $\vg_p\equiv\vg-\vgN$ -- cannot exceed $\eta\az$ in magnitude, where $\eta\sim 1$ depends somewhat on the exact MOND theory (Brada \& Milgrom, 1999). This would be interpreted within the DM paradigm as a pure property of DM halos.
\par
(vii) The central surface density (or equivalently, column density), $\Spz$, of the phantom `dark halos' interpretations of the MOND excess accelerations, is defined as
\beq \Spz=\int_{-\infty}^\infty \r_p(r) dr, \eeqno{mured}
evaluated along the line of sight through the center of the system (for systems that have a well-defined center, such as well-ordered galaxies).
\par
MOND predicts that for systems where the central accelerations are higher than $\az$  (tantamount to the central baryonic surface density $\Sbz>\SM$)
we have
\beq \Spz=p\SM, \eeqno{huta}
with $p\sim 1$, somewhat theory dependent.
This is a pure `DM' property, but the condition for its validity is baryonic.
\par
In the opposite limit, for fully DML systems, in which $\Sbz\ll\SM$, MOND predicts \beq \Spz=s(\Sbz\SM)^{1/2}, \eeqno{liot}
with $s\approx 2$. This is a combined baryon-phantom-DM result.
These laws are derived and discussed in  Milgrom (2009a, 2016).
\par
(viii) For some systems, the mean baryonic surface density, $\bar \Sigma\_B$, cannot much exceed $\SM$.
For disc systems this is because discs with $\bar \Sigma\_B\gg\SM$ (Newtonian discs) are less stable than discs in the MOND regime, having $\bar \Sigma\_B\ll\SM$ (Milgrom, 1989a, Brada \& Milgrom, 1999a).
\par
Nearly-isothermal spheres -- which approximately represent elliptical and other spheroidal galaxies -- also must have $\bar \Sigma\_B\lesssim\SM$, according to MOND (Milgrom 1984).
\par
Interestingly this is a MOND result that pertains to the baryons themselves, without reference to the mass anomalies and putative DM.
\par
To a large extent the above laws also determine approximately the full rotation curves of disc galaxies, given their baryon distribution alone. They tell us that the rotation speed should become constant asymptotically, and what the value of this constant speed should be.
They also tell us that in regions of high acceleration the speed should be Newtonian, and they tell us where the transition should occur.
\par
(ix) The exact prediction of the rotation curve from the baryon mass distribution is somewhat dependent on the specific MOND theory.
However, MOND predicts an approximate formula for the rotation curve, given the acceleration, $\gN$, calculated from Newtonian dynamics in the midplane of the galactic disc (with only baryons). The rotational velocity is determined by the so called `Mass-discrepancy-acceleration relation' (MDAR, Milgrom, 1983, aka RAR, for `radial-acceleration relation')
\beq \frac{V^2(r)}{r}=g(r)=\gN\nu\left(\frac{\gN}{\az}\right), \eeqno{mdar}
where $\nu(y)$ is some MOND interpolating function whose limiting behaviors are dictated by the basic tenets: $\nu(y)\rar 1$ for $y\gg 1$ follows from the `correspondence requirement', and $\nu(y)\approx y^{-1/2}$ for $y\ll 1$ follows from scale invariance and the normalization of $\az$.
\par
A large class of MOND theories predict an exact such MDAR with $\nu(y)$ universal (system independent) (Milgrom, 1994).
In other theories it still holds approximately with the same limiting behaviors.
\par
{\it The MDAR cannot be viewed as equivalent to predicting the rotation curves of individual galaxies.} This is mainly because any observational test of the MDAR involves, perforce, some scatter. There may be important features clearly seen in individual rotation curves that can provide acute tests of MOND and DM, but which may easily be lost in the
 scatter around the MDAR.

\subsection{Laws and theory -- MOND and analogies\label{lawsanalog}}
{\it ``After all these results, ... there is no other decision left for a critic who does not intend to resist the facts, than to award to the quantum of action, which ... has ever-again yielded the same result, ... full citizenship in the system of universal physical constants.''} (Planck, 1920; Nobel-Prize lecture).
\par
An important parallel between MOND and some well-established paradigms, is the `convergence' of phenomenological laws.\footnote{Some of these `laws' may be viewed as `phenomena' -- For example, the photoelectric effect, or the Balmer line spectrum, in quantum theory, time contraction in relativity, or asymptotic flatness of rotation curves in MOND. But, inasmuch as they can be described quantitatively, in mathematical terms, I call them `laws' here.} This occurs especially when seemingly unrelated, or loosely related, phenomena or laws require in their description a new constant of the same dimensions and the same value. Such occurrence points strongly to some unifying, underlying principle or theory.
This is what Planck referred to in the above quote.
\par
Indeed, the eventual emergence of more general quantum theories -- e.g. Heisenberg's matrix mechanics, or Schr\"{o}dinger's wave mechanics, and later, Dirac's relativistic theory -- was preceded by the discovery of several unexpected, experimental laws, and their subsequent explanations in ad-hoc fashions.
These were, in the first instance, the black-body spectrum, with Planck's explanation, and then, e.g., the photoelectric effect, with Einstein's explanation, the Balmer series with Bohr's atomic model, and the temperature dependence of the specific heat of solids, with Einstein's explanation.
\par
These phenomena were not apparently related, and the explanations, all revolving on Planck's constant, $\hbar$, did not add up to a general unifying theory.
Yet, it was highly evident from these alone that there is some such new-physics principle underlying this convergence, as evidenced by the above quote made some years before the advent of an umbrella quantum theory.
\par
The same is true of the many appearances of the `speed of light' constant in relativity:
in Lorentz- and time contraction, in the energy-momentum dispersion relation, and in Schwarzschild's geometry.
\par
As another example,
Merritt (2020) discusses in detail the cogency of such convergence (quoting Perrin on the subject) in connection with the support for the molecular theory brought about by the many appearances of the same value of Avogadro's number in disparate phenomena.
\par
And in MOND? As already said, $\az$ appears in many of the MOND laws. Successful tests of these laws, detailed below in Sec. \ref{testlaws} -- in particular, the accounting for rotation curves of about 200 disc galaxies with a single value of $\az$ (see, e.g., Li \& al, 2018, and references therein). It is found that
\beq \az= 1.2\times 10^{-8}~\cmss,  \eeqno{azero}
with an uncertainty of a few tens of percents.
\par
In MOND, there is an additional aspect to this convergence: As is expanded on in Sec. \ref{signif}, Cosmologically significant accelerations, such as that associated with the accelerated expansion of the Universe, also have values near $\az$.
\par
So, a hundred years ago one could rightly ask: `Why should the black-body spectrum be related to the Balmer series and Bohr's atom, or to the specific heat of solids, if they are not all stemming from one umbrella theory?'
Today one can ask: `Without the umbrella of MOND, why should the $\az$ that enters and determines the asymptotic rotational speed in massive disc galaxies be the same as the $\az$ that enters and determines the mean velocity dispersions in dwarf satellites of the Milky Way and Andromeda galaxies? And why should these be the same $\az$ that enters and determines the dynamics in galaxy groups, which are hundreds of times larger in size and millions of times more massive then the dwarfs (see Sec. \ref{groups} below)?
And why should these appearances in local phenomena in small systems be related to the accelerated expansion of the Universe at large?
\par
More generally, the interplay between phenomenological laws and the Umbrella theory that binds them and shows them as corollaries of a general whole, has been much discussed in the context of the well-established paradigms (e.g., Kuhn, 1962).
I discuss some of these well-known examples here to emphasize that such interplay has similar importance and consequence in the context of MOND.
\par
New paradigms, radically departing from established ones, are almost exclusively induced by some experimental or observational results -- e.g., phenomenological regularities -- that are not satisfactorily explained by the established paradigm.\footnote{Dissatisfaction may be shared by a whole community, or by only a small group who introduce and endorse the new paradigm, as was the case with Copernicus.}
\par
Such phenomenological regularities many time become laws in the new paradigm: The initial observation, which  perhaps pertained to only limited circumstances, or that was not so clear-cut, is elevated to a status of a law of general validity. At the same time, the emerging theory covers much more than the few laws that conduced to it, and among other capabilities it predicts additional laws not discovered or anticipated before.\footnote{It may also happen that a new law is predicted by the theory, but this is not realized before the law is discovered experimentally.}
\par
Kepler's laws, discovered for the solar system, were elevated by Newtonian dynamics -- which was, in turn, largely built on these laws -- to universal laws predicted to apply in all planetary systems. Furthermore, Newtonian dynamics has permitted us to predict the behavior of arbitrary gravitating systems, beyond what is described by specific laws. In the context of planetary motions, for example, it predicted how the Kepler constant (appearing in the 3rd law) would depend on the mass of the central star, and how mutual interactions of the planets themselves affects their motions.
\par
Similarly, the collection of empirical laws concerning electric and magnetic phenomena (Coulomb's, Ampere's, Bio-Savart's, etc.) lead to Maxwell's general theory of electromagnetism.
\par
For MOND, a cornerstone phenomenological law was the asymptotic flatness of rotation curves.
At the time of MOND's advent, such asymptotic flatness already appeared as a tentative property of disc galaxies. Looking e.g., at Figs. 5 and 6 of Rubin \& al., 1980, or at Fig. 3 of Bosma, 1981 (also published in his PhD thesis, Bosma, 1978), we see a clear trend for asymptotic flatness, but not for all galaxies, and flatness is not exact.\footnote{Rubin \& al., 1980 state that {\it `Most galaxies exhibit rising rotational velocities at the last measured velocity.'}}
\par
Indeed, while the rotational speeds known at the time were incontestably too high to be compatible with Newtonian dynamics, strict, universal, asymptotic flatness was not generally established, nor accepted. For example, the late John Bahcall wrote me on February 26 1982, commenting on my initial MOND trilogy: {\it `...There are conventional astronomical explanations for all the facts. This is particularly clear in the case of the ``flat'' rotation curves, which appears to be a crucial example in your thinking. I don't think much can be said within the uncertainties about how flat rotation curves are at large distances (outside the Holmberg radius). Fifteen percent changes may occur and be obscured by observational errors, geometrical configurations, non-circular motions and other factors. I have recently become convinced of this fact in my study of the dynamical implications of ``apparently flat'' rotation curves...'}
\par
And, from another astrophysicist's comments from July 28 1983:
{\it `...I'm impressed by the consistency of your proposal with the data, but curious whether the continued rise in $V^2$ with $r$ at large $r$ reported by Rubin et al. 1982 doesn't pose problems.'}
\par
Even today, asymptotic flatness is not a law recognized by DM advocates, since the purported DM halos are of finite mass, which implies asymptotically falling rotation curves. In MOND it is a physical law (for isolated bodies).
\par
Still, this tentative asymptotic flatness was the main reason for my looking for alternatives to DM. I made it the seminal axiom of MOND, ignoring the uncertainties, inexactitudes, and departures from flatness -- which indeed Bahcall was right about --  in order to have some principle to found the alternative on.
\par
And, since asymptotic dynamics is one instance of low accelerations, I generalized and constructed MOND on departure from Newtonian dynamics always when accelerations are low. This, in turn, has lead to the introduction of $\az$, and to predictions of the many further laws, and then to complete theories.
\subsection{Significance of $\az$\label{signif}}
It was noted right at the advent of MOND, and with increasing focus over the years, that $\az$ is near in value to some accelerations of cosmological significance (detailed account of this can be found in Milgrom, 2020). At first, the connection was noted (Milgrom 1983) with $H_0$, the present expansion rate of the Universe -- or the Hubble-Lema\^{i}tre constant. Then, the connection was  made with the accelerated expansion of the Universe (associated with a cosmological constant, or `dark energy'), at first as only a hypothetical entity (Milgrom 1989); then, with more concreteness (Milgrom 1994), after the first hints of a cosmological constant emerged (Efstathiou \& al. 1990); then (Milgrom, 1999) with even more purpose, after a cosmological-constant-like effect in cosmology was definitely identified.
\par
These connections can be summarized by the near equalities
\beq \baz\equiv 2\pi\az\approx c H_0~~~~~~~~~~~~~\baz\approx c^2(\Lambda/3)^{1/2},  \eeqno{coinc}
where $c$ is the speed of light, and $\Lambda$ is the cosmological constant, taken here to have dimensions of length$^{-2}$
($\Lambda$ relates to the density of `dark energy').\footnote{The fact that these two seemingly unrelated cosmological parameters are similar in value, $c H_0 \approx c^2(\Lambda/3)^{1/2}$, if not a mere coincidence, might be an important clue.
It is the same statement, within GR, that today the `dark energy' density is of the same order as that of matter.}
\par
The `coincidences' (\ref{coinc}) can be cast in other useful forms: Define the MOND length as
\beq \lM\equiv c^2/\az. \eeqno{mondlength}
Then, relations (\ref{coinc}) tell us that $\lM$ is of the order of the size of the presently observable Universe:
\beq \lM\approx  \lU, \eeqno{length}
Where $\lU$ can be the `Hubble distance', $c/H_0$, or the `de Sitter radius' associated with the cosmological constant, $\dsr\equiv(\Lambda/3)^{-1/2}$.
\par
We can also cast relations (\ref{coinc}) in terms of the MOND mass,
\beq M\_M\equiv c^4/\azg  \eeqno{mondmass}
as $M\_M\approx M\_U$, where $M\_U$ is the total `mass' (including `dark energy') within the observable Universe.
\par
These relations may be hinting at the underlying, more fundamental MOND theory.
Indeed, there have been many proposals for how to account for such a connection in the frameworks of various
`microscopic' pictures for MOND
(e.g., Milgrom 1999, Pikhitsa, 2010, Ho, \& al., 2010, Kiselev \& Timofeev, 2011, Klinkhamer \& Kopp, 2011, Li \& Chang, 2011,
Pazy \& Argaman, 2012, Verlinde, 2017, Milgrom, 2019a).
\par
One intriguing possibility (Milgrom, 1999) builds on the following observation:
There are various physical phenomena that connect an acceleration $a$ with some length given by $\ell_a\equiv c^2/a$. For example, this is the typical wavelength of the so called `Unruh radiation' associated with such an acceleration. For an accelerating charge, $\ell_a$ is the distance where the radiation field begins to dominate the near field, etc.
It may be said that a system accelerating at $a$ probes, in some regards, to a distance $\ell_a$.
Then, local MOND dynamics could somehow result from the fact that bodies with $a\gg\az$ probe distances $\ell_a\ll\lM\approx\lU$; so they are not aware of the curved nature of the Universe. But, systems with $a\ll\az$ and hence $\ell_a\gg\lU$, do probe to the cosmological horizon, sense the curved nature of space time, and so obey different dynamics. What we still lack is a mechanism that actually makes the curved-space-time Universe affect local dynamics (heuristic suggestions are discussed in Milgrom, 1999 and the subsequent works listed above).
\par
In the well-known quantum analogy, there is a relation between momentum $P$ and a distance $\ell\_P\equiv\hbar/P$, introduced, e.g., through the uncertainty principle.
For example, the quantum dynamics of a particle confined to a box of size $\ell\_B$ (analogous to the `size' $\lU$) depends on its momentum in relation to $P\_B\equiv \hbar/\ell\_B$ ($P\_B$ is analogous to $\az\approx c^2/\lU$).
Particles with high momentum, $P\gg P\_B$, are practically oblivious to the confining box, and behave almost like free (quantum) particles.\footnote{In the sense that the spectrum at these high momenta is almost continuous, with energy and momentum relative differences between neighboring levels being very small.} But the spectrum of particles with $P\sim P\_B$ clearly shows that they are confined in a box of size $\ell\_B$.
And thus $P\_B$ becomes an effective `boundary constant' for this boxed system.
\par
The MOND length, $\lM$, MOND mass, $M\_M$, and a MOND time, $t\_M\equiv c/\az$, may, in themselves, have physical significance, and might be physically the more fundamental, and more indicative of the origin of MOND, than $\az$ itself.
Analogously, in quantum theory one combines $\hbar$ with c and $G$ to form the Planck length, time, and mass
\beq \ell\_P\equiv \left(\frac{\hbar G}{c^3}  \right)^{1/2},~~~t\_P\equiv \frac{\ell\_P}{c},~~~M\_P\equiv \frac{\hbar}{c\ell\_P}.   \eeqno{planck}
These quantities are indicative in different ways of where quantum gravity may be important. For example, a black hole having a mass near $\M\_P$ requires quantum gravity for its description.
\par
There is a historical example of a numerical coincidence between seemingly unrelated quantities  pointing to deeper connections: Weber and Kohlrausch measured and published in 1856 the so called `ratio of electromagnetic and electrostatic units', call it $v\_e$.
This is a constant with the dimensions of velocity that enters a combination of the disparate phenomenological laws of electromagnetism known at the time.\footnote{It can be written as $v\_e=1/\sqrt{\epsilon\_0\mu\_0}$ with $\epsilon\_0$ and $\mu\_0$ the permittivity and permeability of the vacuum.} They noted that their result for $v\_e$ is near in value to the then-known value of the speed of light (in its literal sense). (Maxwell, 1892 gives a detailed account of this.) Their measurement method had nothing to do with light or its speed. Weber and Kohlrausch seem not to have made much of this `coincidence', and it was left to others (e.g., Maxwell) to make this epochal observation that light is an electromagnetic phenomenon -- a warning to mind our `coincidences'.
\par
It is part of the craft of a scientist to decide which `coincidences' to take seriously and which not to.
\par
I say more on this important issue of the coincidences in Sec. \ref{cosmology}.
\par
Relations (\ref{coinc}) have also some `practical' implications.
To give but one example, they imply that we cannot have a system that involves highly relativistic gravity, and is in the DML
(except the Universe at large, which is marginally in the MOND regime).
For example, we cannot have deep-MOND  black holes: Gravity is `relativistic' at a distance $r\sim MG/c^2$ from a mass $M$. To also be in the DML at this radius means that $MG/r^2\ll\az$, which together imply $r\gg \lM\approx \lU$; so $r$ has to far exceed the `size of the Universe'.
\subsection{MOND theories\label{theory}}
While many major predictions of MOND follow from only its basic tenets, clearly, we need to dress this skeleton of axioms with a full-fledged MOND theory that would allow us to calculate general systems, and account for details of the phenomena. MOND theories were reviewed in Milgrom (2014, 2015); here I discuss only some generalities concerning them.
\par
Several relativistic MOND theories have been proposed and studied to date. A notable landmark was the advent of Bekenstein's Tensor-Vector-Scalar theory (TeVeS) (Bekenstein, 2004), which was built on ideas from Sanders, 1997. TeVeS was the first relativistic MOND theory with realistic gravitational lensing; and it was the bellwether for more relativistic MOND versions.
\par
In particular, Skordis \& Z{\l}o\'{s}nik, 2019 have recently advanced a promising class of extensions of TeVeS. In common with TeVeS, this class of relativistic MOND theories describe gravity by a metric tensor (as in GR) with auxiliary vector and scalar fields. The attractive aspect of this class is that it predicts that the tensor gravitational waves propagate with the same speed as light, under all circumstances.\footnote{Bekenstein's original version of TeVeS does not pass this test, and has some other deficiencies. But the work of Skordis \& Z{\l}o\'{s}nik, 2019 keeps TeVeS alive.} This negotiates an important hurdle posed by the recent observation
of electromagnetic and (tensor) gravitational-wave signals arriving at the same time from an event of a double-neutron-star merger.\footnote{Gravitational waves of the vector and scalar fields should be immaterial in this context. The gravitational acceleration both near the emitting source (double neutron stars or black holes), and the detectors (the inner solar system) are many orders of magnitude larger than $\az$. Thus, MOND predicts that GR prevails there to a high accuracy, and it allows only the tensor waves. Vector and scalar waves are practically not produced, and could not, in any event, reach the detectors on Earth.}
\par
The MOND theories proposed to date satisfy the basic tenets of MOND (by definition) and thus make all the salient predictions discussed above in Sec. \ref{laws}. They differ in important details; but this has not yet led to clear observational discrimination between them, because the clear-cut and easy-to-interpret observations concern those predictions that follow from basic MOND tenets, and are shared by all these theories.
\par
Standard dynamics are underlaid by several basic principles, such as locality,\footnote{The notion that physical phenomena can be described as a collection of events each occurring at one space-time point.} Lorentz invariance, and general covariance. These dynamics have been amply tested in various circumstances. However, they do fail in the realm of the galaxies without invoking DM, with
MOND highlighting the fact that this occurs consistently for accelerations below $\az$.
So, in the least, MOND points to the fact that standard dynamics and their principles have not been verified in the region of low accelerations.
 \par
Thus, in attempting to generalize standard dynamics to include the MOND tenets, it is not clear what general principles obeyed by standard dynamics should be retained in the full MOND theory.
\par
One is naturally loath to abandon long-held principles.
But, taking a lesson from quantum theory and relativity, we know that major, well-tested, long-held principles of classical Newtonian dynamics have had to be abandoned in the new paradigms. In Quantum theory, such are the continuity of physical states, which gave way to discreteness (e.g., of the values the energy of an atom can take), and the ability to predict exactly the outcome of a measurement, replaced by the ability to predict only probabilities.\footnote{The continual attempts to establish quantum theory on deterministic `hidden variable' schemes witness the difficulty some had palating these rejections of long-held and `intuitive' principles. Forgoing the basic underlying circular motions of planetary orbits to be replaced by elliptical orbits may have been as painful.}
In the case of relativity, the abandonment of the concept of absolute time is an example.
\par
So it may well be with MOND, that some underlying principles of GR may have to be abandoned in the DML.
There are already some proposals in this vein, assuming, for example breakdown of local Lorentz invariance in the DML (Blanchet \& Marsat, 2011; Sanders, 2011), of locality (e.g., Deffayet \& al., 2014), or general covariance (Milgrom, 2019b).
\par
Another crucial point is that MOND as we know it now is, arguably, only
an approximate `effective field theory' that approximates some more fundamental scheme at a deeper stratum -- some `FUNDAMOND' -- conceptually, in a similar way to thermodynamics being an approximation of the statistical-mechanics, microscopic description. In principle, one of the existing, relativistic MOND theories may turn out to be the final word -- the FUNDAMOND. However, for reason discussed below, I feel that this is not likely; namely, that all the MOND theories we now have are, probably, at best some useful approximations of such a FUNDAMOND. Useful indeed they are because they allow us to make general computations (e.g. of galaxy formation, evolution, and interactions) in a theory that conforms with the basic MOND tenets, that obeys all the standard conservation laws (of momentum, energy, and angular momentum), etc. But one is not sure how far we can actually trust these theories.
\par
Even well-working theories may leave one with a sense that they can be only approximation of a deeper theory. This may be a result of perceived conceptual or phenomenological shortcomings of the theory.
For example, some have considered the probabilistic aspects of quantum theory as unsatisfying; so they labour to derive it as an approximate description of some deeper-level.
General relativity is thought to be a classical (low-energy) approximation of some fundamental `quantum-gravity' theory that we still lack.
The standard model of particle physics, which works very well in describing many phenomena, is thought to be an approximate, effective theory, because it has many parameters whose values have to be determined by experiment, because it does not describe gravity, cannot account for finite neutrino masses, does not account for the dominance of matter over antimatter in the Universe, and, for those who believe in DM, it has no room for it.
\par
In the case of MOND, the urge to look for a deeper-theory origin is twofold: a. The above mentioned `coincidence' (\ref{coinc}) might be a hint that MOND will have to be understood within a larger frame that includes the description of cosmology. b. All MOND theories proposed to date introduce by hand the interpolation between standard dynamics and the DML as the acceleration goes from high to low. Clearly, we need an underlying theory in which such interpolation will emerge from the theory itself.
\par
To appreciate this last point better, note that in laws of quantum theory and relativity there also appear interpolating functions between the `classical' and the `new' regimes.
The black-body spectrum was the first quantum-mechanical `interpolating function', which induced Planck to introduce his constant. It interpolates between the Rayleigh-Jeans, classically-calculated part of the spectrum (attained when $\hbar\rar 0$), and the Wien, fully-quantum region. Other such interpolating functions are the specific heat of solids as a function of temperature, interpolating between the classical, Dulong-Petit expression and the quantum branch. The expression for barrier penetration probability is another, and so forth.
\par
In relativity, the dependence of inertial mass on velocity $m=m_0/\sqrt{1-v^2/c^2}$ is an example; another is the dependence on velocities of the force between moving charges.
\par
But these interpolating functions do not appear in the basic formulation of the theory. Furthermore, they are different for different laws and phenomena. What is common to them is that they reduce to the classical result in the appropriate limit ($\hbar\rar 0$, or $c\rar\infty$).
This is also expected to be the situation in an eventual underlying theory of MOND.

\subsection{Tests of the  MOND laws \label{testlaws}}
Here I describe briefly observational tests of some of the MOND laws listed in Sec. \ref{examples}.
Some of these tests were performed on disc galaxies, where the main tool involves the circular motions of constituents in the galactic disc.
\par
Unlike the discs of galaxies, where gravity is balanced by ordered rotational motions, in so called `pressure-supported systems' -- such as globular clusters, elliptical galaxies, dwarf spheroidal satellite of galaxies, galaxy groups, and galaxy clusters -- the `inertial forces' that balance gravity are associated with `random' motion, analogous to pressure.
\par
At present, the main tool for testing such systems, inasmuch as they are in the DML -- is the virial relation of law (iv), eq.(\ref{msigma}).

\subsubsection{Asymptotic constancy of orbital velocity: $V(r)\rar V\_{\infty}$}
As described in Sec. \ref{lawsanalog}, this law was already known to hold approximately at the time of the advent of MOND, and was the main axiom in the construction of MOND, which elevated it to a law of Nature.
\par
\begin{figure}[!ht]
\includegraphics[width=1.\columnwidth]{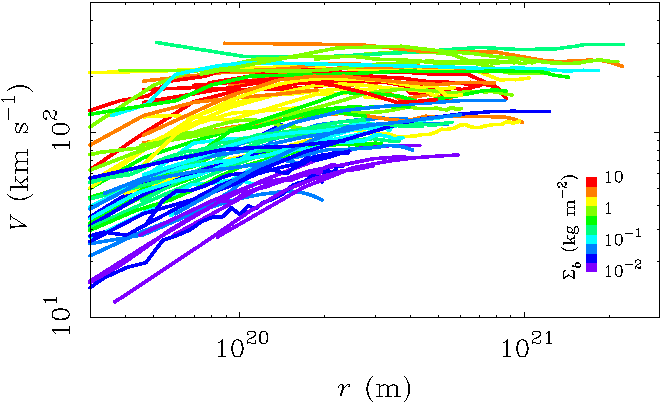}
\caption{A compilation of many rotation curves of disc galaxies. The colors mark the surface density of the galaxy, which is a measure of the characteristic Newtonian gravitational acceleration in the disc produced by baryons (Stacy McGaugh, private communication). \label{flatrcs}}
\end{figure}

Figure \ref{flatrcs} shows a compilation of many rotation curves -- rotational speed as a function of radius. We can see the tending of the speed to become radius-independent at large radii. See also some separate rotation curves below in Sec. \ref{rcindiv}.
\subsubsection{Mass-asymptotic-speed relation}
\begin{figure}[!ht]
\includegraphics[width=2.3\columnwidth]{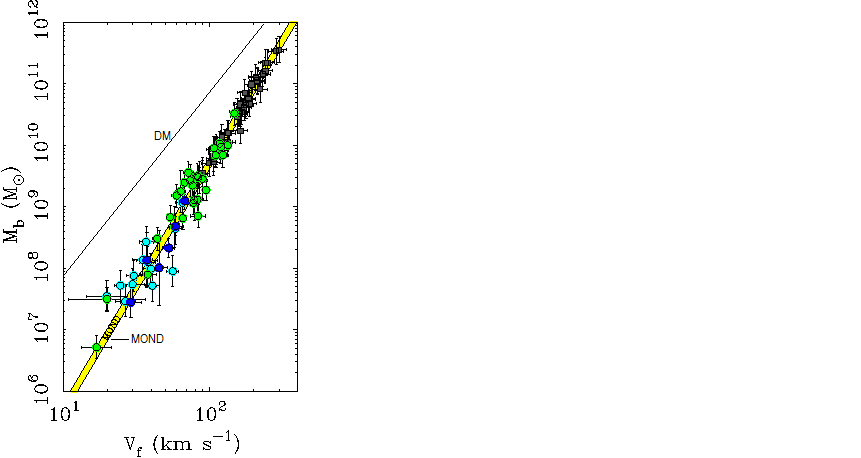}
\caption{Adapted from McGaugh (2012). The total observed, baryonic mass of each galaxy, $M_b$, is plotted against the rotational speed at the flat part of its rotation curve, $V_f$.  From the original caption: `The band shows the expectation of MOND (Milgrom 1983). The width of the band represents the $\pm 1\s$ uncertainty in $\az$ based on detailed fits to the rotation curves of a subset (Begeman \& al., 1991) of the star dominated galaxies (gray squares).'  The thin line marked `DM' is the expectation from \LCDM if the cosmic ratio of baryons to DM is retained in galaxies. It is based on the relation between halo mass and maximum rotational speed of the halo as gotten from \LCDM simulations, with the former reduced by the cosmic fraction (this line is from another Figure in McGaugh, 2012).\label{btfr}}
\end{figure}

Figure \ref{btfr} shows one of many tests of the MASR, law (iii) (e.g., Sanders, 1996; Noordermeer \& Verheijen, 2007; McGaugh, 2011; McGaugh, 2012; Papastergis \& al., 2016).  The baryonic mass of many galaxies is plotted against the asymptotic speed of their rotation curves. The prediction of law (iii), as per eq. (\ref{masr}), is also shown.
\par
\begin{figure}[!ht]
\begin{tabular}{ll}
\includegraphics[width=0.9\columnwidth]{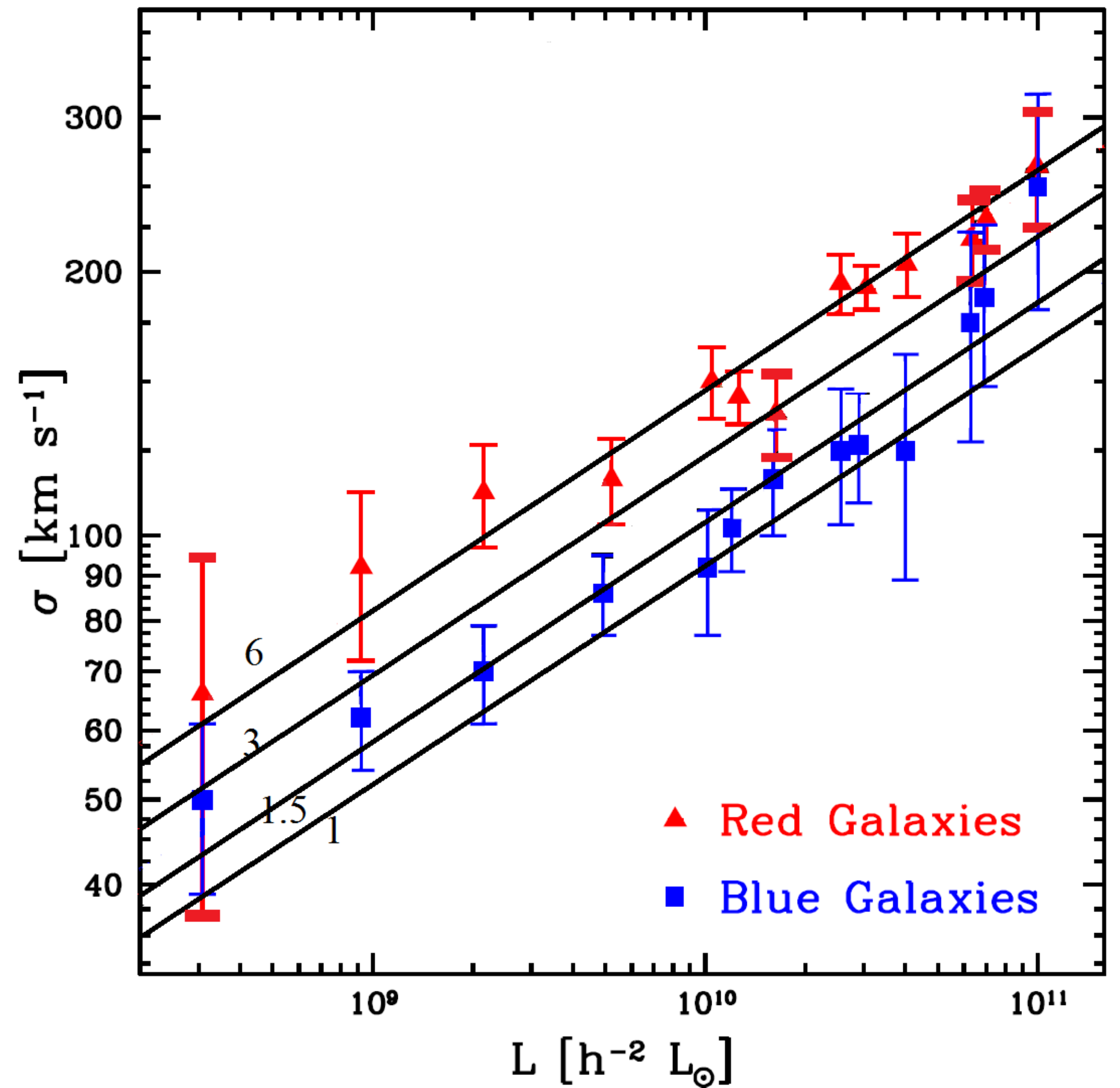}
\end{tabular}
\caption{An equivalent of the asymptotic rotational speed (called here $\s$) plotted vs. the galaxy luminosity (a proxy for its baryonic mass) based on statistical analysis of galaxy-galaxy lensing. Each data point represent a bin of all lens galaxies within a certain luminosity range. Blue squares represent `late type', disc galaxies, and red triangles `early type', elliptical galaxies. The points with error bars are the result of lensing analysis by Brimioulle \& al., 2013. The lines show the predictions of the MOND MASR for different mass-to-light ratios of the galaxies, which are appropriate for the two types of galaxies shown (from Milgrom, 2013).\label{lensing}}
\end{figure}

A test of the MASR using analysis of galaxy-galaxy weak lensing (instead of rotation curves), which probes a very large number of galaxies (albeit only statistically), comprising not only disc galaxies, but galaxies of all types, is shown in Fig. \ref{lensing}, based on results of  Brimioulle \& al. (2013). This implicitly incorporates law (ii), and is thus also a test of the MOND prediction eq. (\ref{bending}). Similar analysis of a different data set was done by Brouwer \& al., 2017, with similar results.

\subsubsection{Mass-discrepancy-acceleration relation in disc galaxies}
\begin{figure}[!ht]
\includegraphics[width=1.\columnwidth]{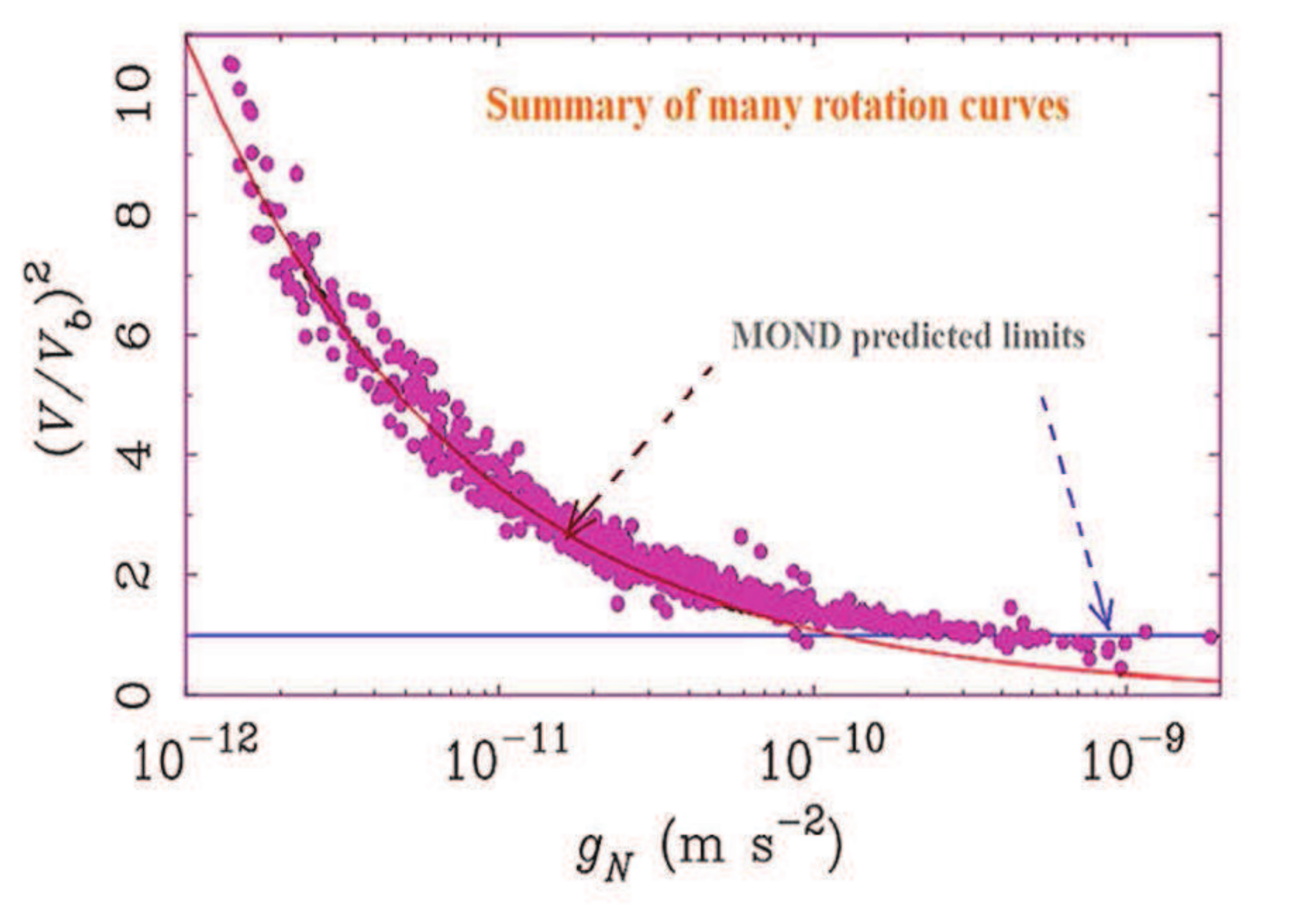}
\caption{A test of the MDAR, eq. (\ref{mdar}): A plot of $g/g_N$ vs. $g_N$ at different radii for 73 disc galaxies  (McGaugh, 2015, private communication). The prediction is that all points fall on the line $\nu(\gN/\az)$ as in eq. (\ref{mdar}). Also shown are the two MOND-predicted asymptotes of this relation (the low-acceleration asymptote for $\az= 1.2\times 10^{-8}~\cmss$; note that the units in the plot are $\mss$).\label{mdar1}}
\end{figure}

Over the years, there have been quite a few tests of the MOND MDAR [eq. (\ref{mdar})], with ever improving data (larger galaxy samples, better measurements of rotation curves, and better determination of the baryon mass distribution). The MDAR has been plotted, e.g., by Sanders, 1990; McGaugh, 1999; Tiret \& Combes, 2009; Wu \& Kroupa, 2015; and McGaugh \& al., 2016.
Figs. \ref{mdar1} shows one of these.
Each disc galaxy contributes many points in the plot, coming from different radii in the galaxy. Some galaxies, which are of low-acceleration at all radii, contribute only to the low-acceleration asymptote, and some contribute to both sides of the transition.
\par
Note, importantly, that the confirmation of the MDAR vindicates several of the MOND laws, in which $\az$ appears in three {\it independent} roles:
a. Law (v): the tightness of the MDAR -- which is `consistent with zero intrinsic scatter' -- and, in particular, it being so tight in the transition region shows that indeed the break from Newtonian dynamics always occurs at the same value of the acceleration. The value of $\az$ in this role of `boundary constant' can be read, e.g., from the intersection of the two asymptotes, somewhat above $1\times 10^{-8}~\cmss$ (note that the units in the plot are $\mss$).
b. Laws (i) and (iii): If the asymptotic rotational speed is constant (occurring where $g\ll\az$) -- as in law (i) -- and given by $V\_\infty=(MG\az)^{1/4}$ (law (iii), then at all the radii, $r$, where this velocity is attained, $g=V^2\_\infty/r=(MG\az)^{1/2}/r$. Since at these radii the galaxy may be considered a point mass, the Newtonian acceleration there is $\gN=MG/r^2$. These together give the unique relation $g=(\gN\az)^{1/2}$. Many points on the low-acceleration branch of the MDAR come from such regions, and the fact that they fall around the MOND asymptote both confirms the above laws and gives an independent determination of $\az$ at the canonical value.
c. Many points on the low-acceleration branch come not from asymptotic regions but from within galaxies with low-accelerations everywhere. These constitute a completely independent test, as there is no known reason why such low-acceleration interior regions should have any relation to the asymptotic regions in the far outskirts of galaxies. It is only the low accelerations that they have in common. These points thus also give us a value of $\az$ consistent with all other determinations.

\subsubsection{MDAR for pressure-supported systems\label{masrpress}}
\begin{figure}[!ht]
\includegraphics[width=1.\columnwidth]{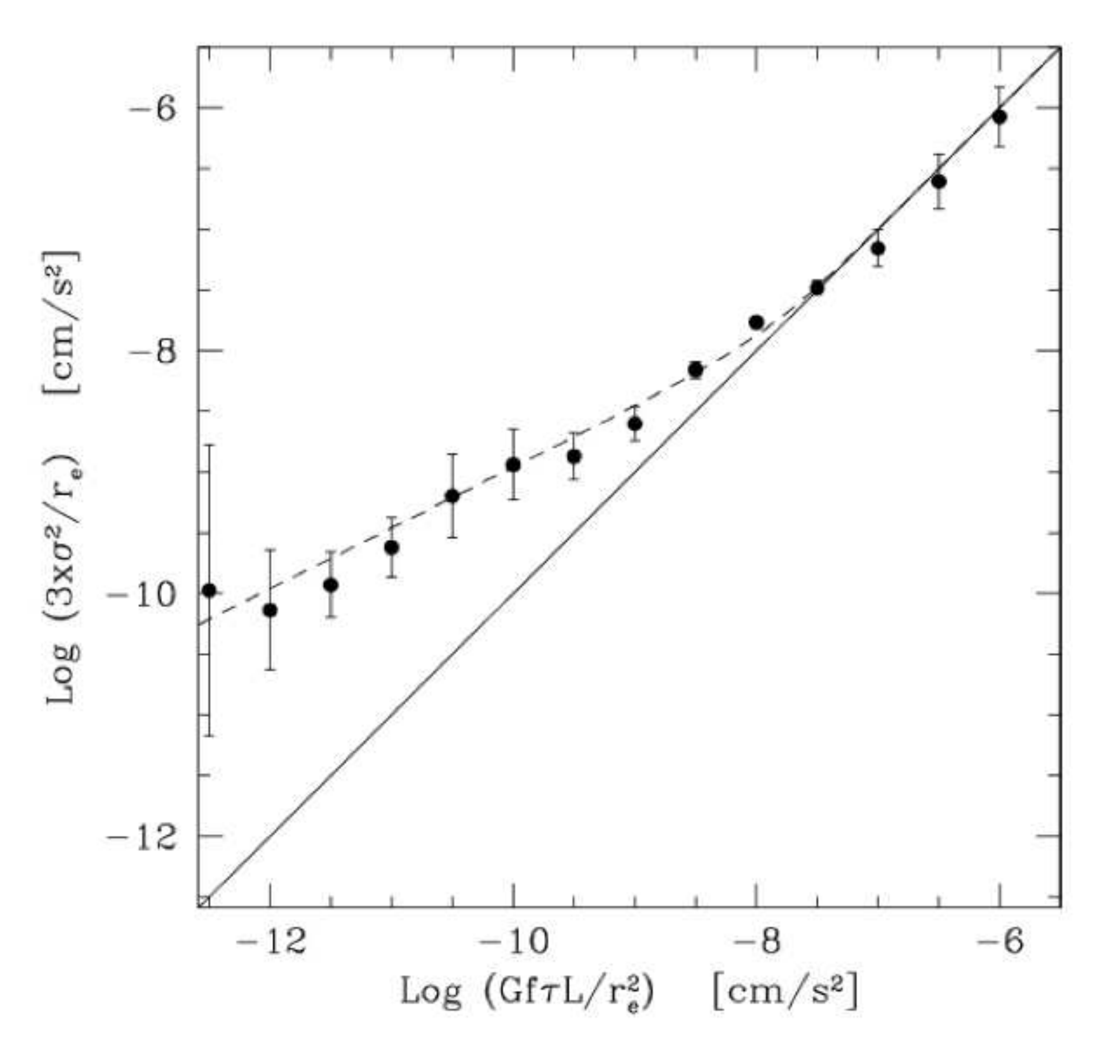}
\caption{The MDAR [eq. (\ref{mdar})], here as a plot of  $g$ vs. $g_N$ (or rather of parameters that represent these quantities) for different types of pressure-supported systems, from  Scarpa, 2006. The Solid line is the Newtonian expectation with no DM. The dashed line is the MOND prediction for the canonical value of $\az= 1.2\times 10^{-8}~\cmss$.  \label{scarpa}}
\end{figure}

Figure \ref{scarpa} shows a compilation from Scarpa, 2006, showing the analog of the MDAR, eq. (\ref{mdar}), for many types of pressure-supported systems.
\subsubsection{Maximal `halo' acceleration}
The prediction of law (vi), pointed out in Brada \& Milgrom, 1999, was first tested in Milgrom \& Sanders, 2005.
More recently, Tian \& Ko, 2019 deduced the excess accelerations (dynamical minus baryonic) -- which in the DM paradigm are interpreted as produced by the DM halo -- using the disc galaxies from the large SPARC sample (see below) and from additional elliptical galaxies. They find (e.g., their Fig. 1) that indeed the (phantom) halo accelerations plotted against the baryonic one shows a clear maximum of about $\az$ as MOND predicts.

\subsubsection{Relations between Dynamical, Baryonic, and phantom-DM central Surface densities}
\begin{figure}[!ht]
\includegraphics[width=1.\columnwidth]{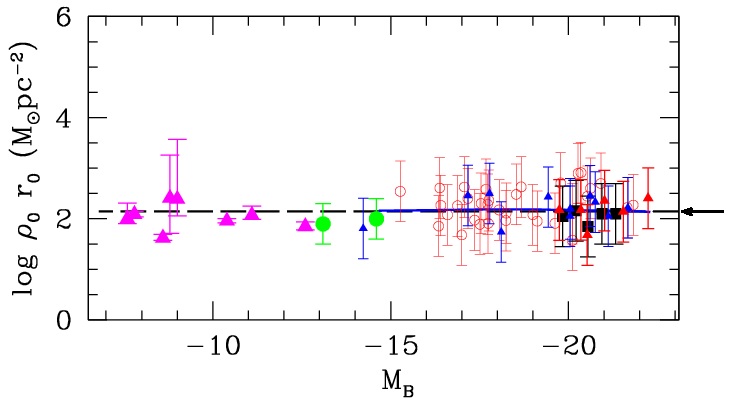}
\caption{Central column densities of best-fitted DM halos of disc galaxies from Donato \& al., 2009, analyzed and discussed in Milgrom, 2009a. The arrow at the right shows the value of $\SM$, predicted by MOND (for the canonical value of $\az= 1.2\times 10^{-8}~\cmss$).\label{donato}}
\end{figure}

\begin{figure}[!ht]
\includegraphics[width=0.9\columnwidth]{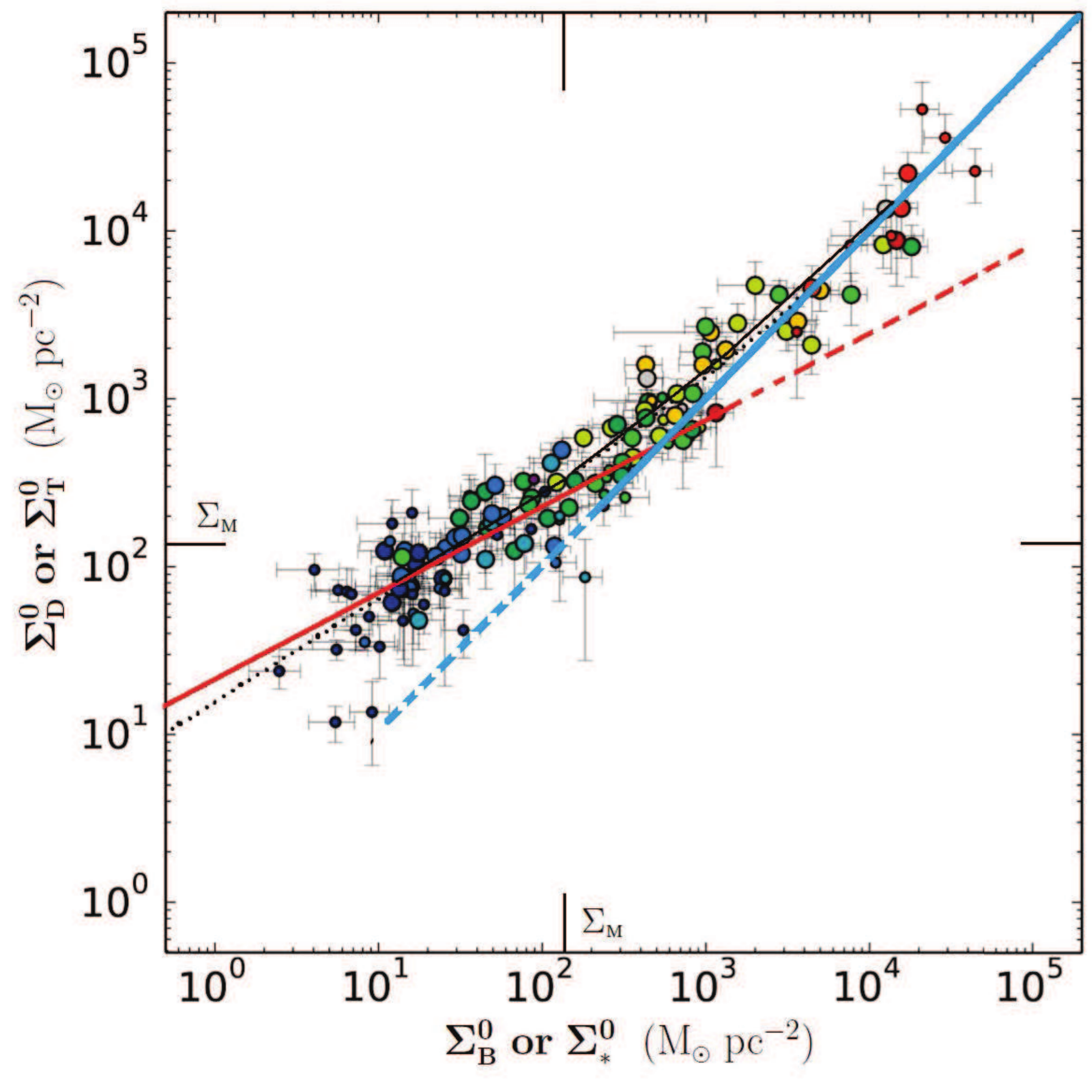}
\caption{The total central column density plotted against the baryonic value (from Milgrom, 2016). The data are from Lelli \& al., 2016, and the dotted line is their (not physically motivated) best 3-parameter fit to the data. The black continuous line is the MOND parameter-free prediction (not a fit) derived in Milgrom, 2016 for the canonical value of $\az= 1.2\times 10^{-8}~\cmss$. The red and blue (partly solid partly dashed) lines are the predicted MOND asymptotes. The value of the MOND surface density, $\SM$, is marked on the axes. According to Lelli \& al.: `The observed scatter is small and largely driven by observational uncertainties...(this gives a) central density relation with virtually {\it no} intrinsic scatter'.\label{sigsig}}
\end{figure}

As a test of law (vii),
Fig. \ref{donato} shows the central surface densities of best-fitted `DM halos' of disc galaxies from Donato \& al., 2009, with the predicted MOND value $\Spz=\SM$ marked.
This confirms the first part of law (vii), eq. (\ref{huta}).
Figure \ref{sigsig} shows a test of the more detailed MOND prediction of a relation between the central surface densities -- baryonic and dynamical. The latter represents the sum of the baryonic and the putative phantom DM needed to explain the dynamics within Newtonian dynamics. Using the large sample of SPARC disc galaxies, Lelli \& al., 2016
found a tight relation between the two quantities. It follows quite closely the MOND prediction -- which involves no free parameters -- derived afterwards in Milgrom, 2016.

\subsubsection{Rotation Curves of Disc Galaxies\label{rcindiv}}
Rotation-curve analysis is the flagship of MOND predictions and testing, and I dwell on it here in more detail. It is also an opportunity to recall some history of the subject.
\par
\begin{figure}[!ht]
\includegraphics[width=0.9\columnwidth]{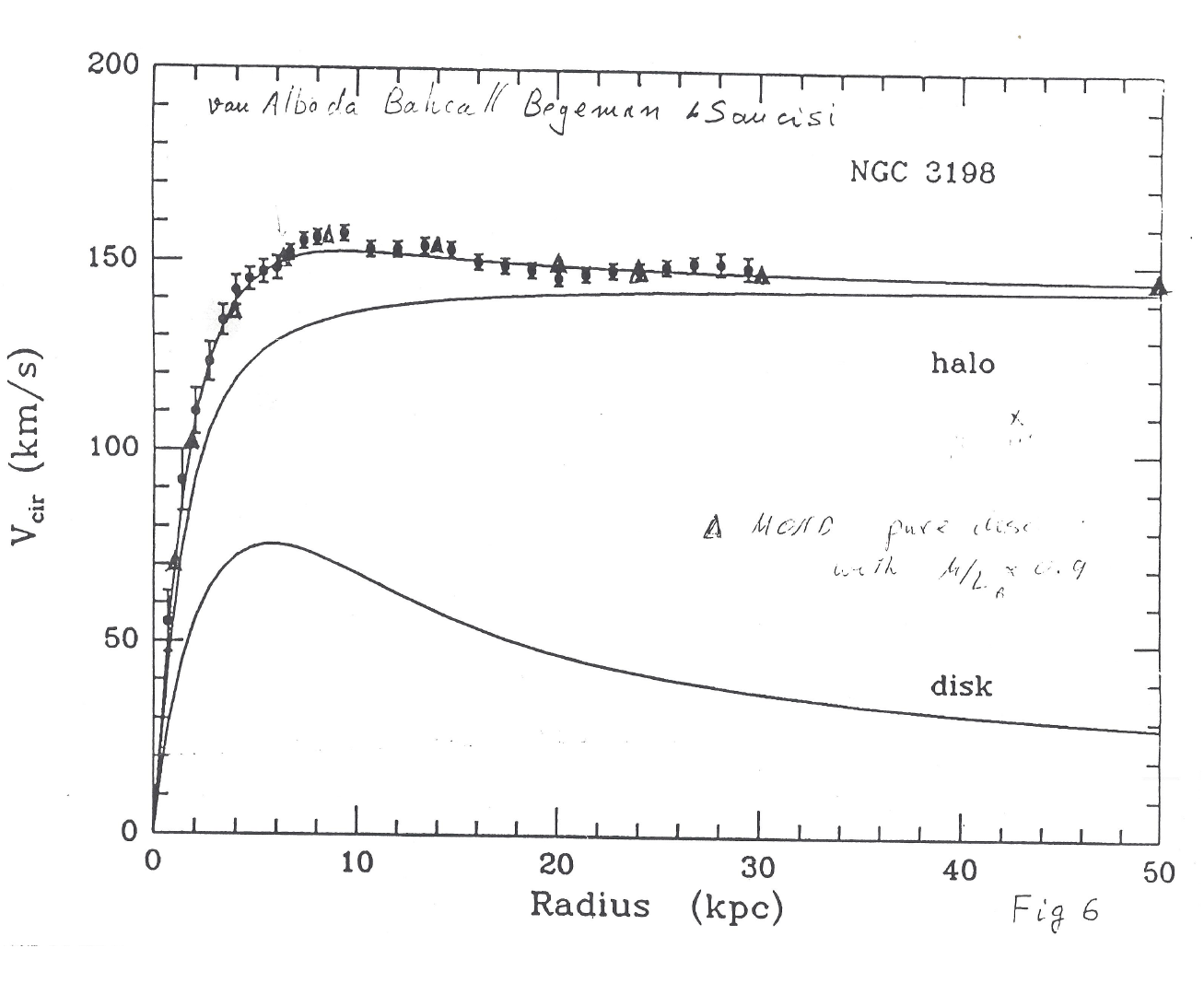}
\caption{The very first calculation of a rotation curve with MOND (for the disc galaxy NGC 3198), taken from my notes of June 1984. The data are from a preprint by van Albada et al. given me by John Bahcall (later published in van Albada et al., 1985) with their model fits for the Newtonian curves for the baryonic disc and the dark matter halo, and the total best fit which involved several parameters. My MOND points for the fit with a single parameter -- the mass-to-light ratio for the disc -- are the hand-drawn triangles for the rotation speeds at several radii.\label{firstRC}}
\end{figure}

The very first MOND analysis of a rotation curve (of the disc galaxy NGC 3198) is shown in Fig. \ref{firstRC} for historical interest. It is taken from my (unpublished) June 1984 notes.
This somewhat tentative result\footnote{The value of $\az$ used and the distance to the galaxy were off the now accepted values, hence the too-low value of the mass-to-light ratio (there is here a degeneracy among the three parameters).} (together with two others) was published only a few years later, in Milgrom \& Bekenstein, 1987, in the proceedings of an IAU symposium held in 1985.
\par
\begin{figure}[!ht]
\includegraphics[width=0.9\columnwidth]{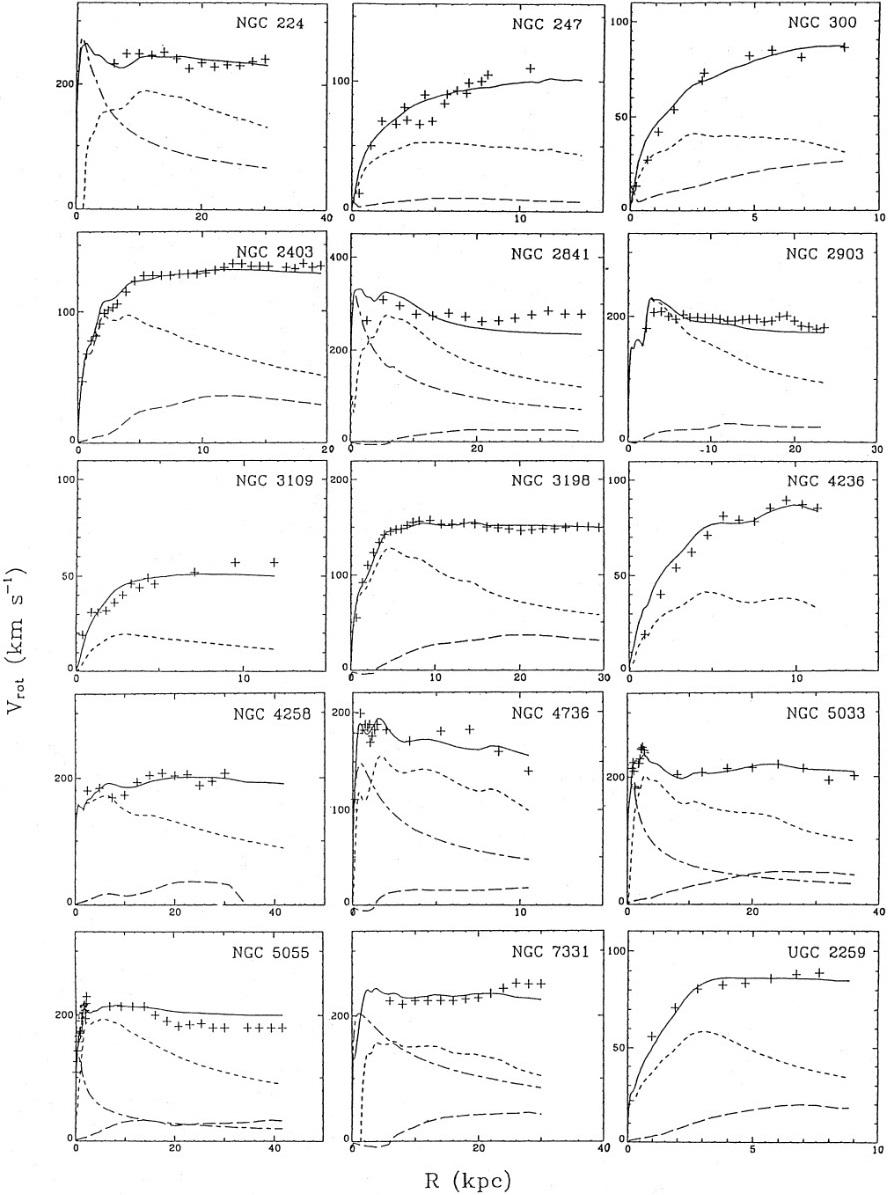}
\caption{The first extensive rotation-curve analysis in MOND done by Steve Kent, and published in Milgrom, 1988.
From the original caption: `Revised MOND curves calculated by Kent (private communication) (solid lines) as compared with the measured curves (plus signs). A single value of the acceleration parameter is used for all model curves. The mass models used include contributions from the stellar disc (short-dashed lines) and the bulge (the long-short-dashed lines), as well as from the observed hydrogen (long-dashed lines).'\label{kent}}
\end{figure}

The first extensive rotation-curve analysis in MOND used a sample of 15 disc galaxies. It was performed by Steve Kent, and published in Milgrom, 1988. Its results are shown in Fig. \ref{kent}, also for historical and pedagogical purposes. The value of $\az=1.3\times 10^{-8}~\cmss$ that was used for all galaxies is very near the presently accepted canonical value given above, and emerged already from this early analysis as best fitting the sample as a whole.

\par
Many more such analyses followed, a few of which are as follows: Begeman \& al., 1991; Sanders, 1996; de Blok \& McGaugh, 1998;  Barnes \& al., 2007; Gentile \& al., 2011; Hees \& al., 2016.
\par
\begin{figure}[!ht]
\includegraphics[width=.9\columnwidth]{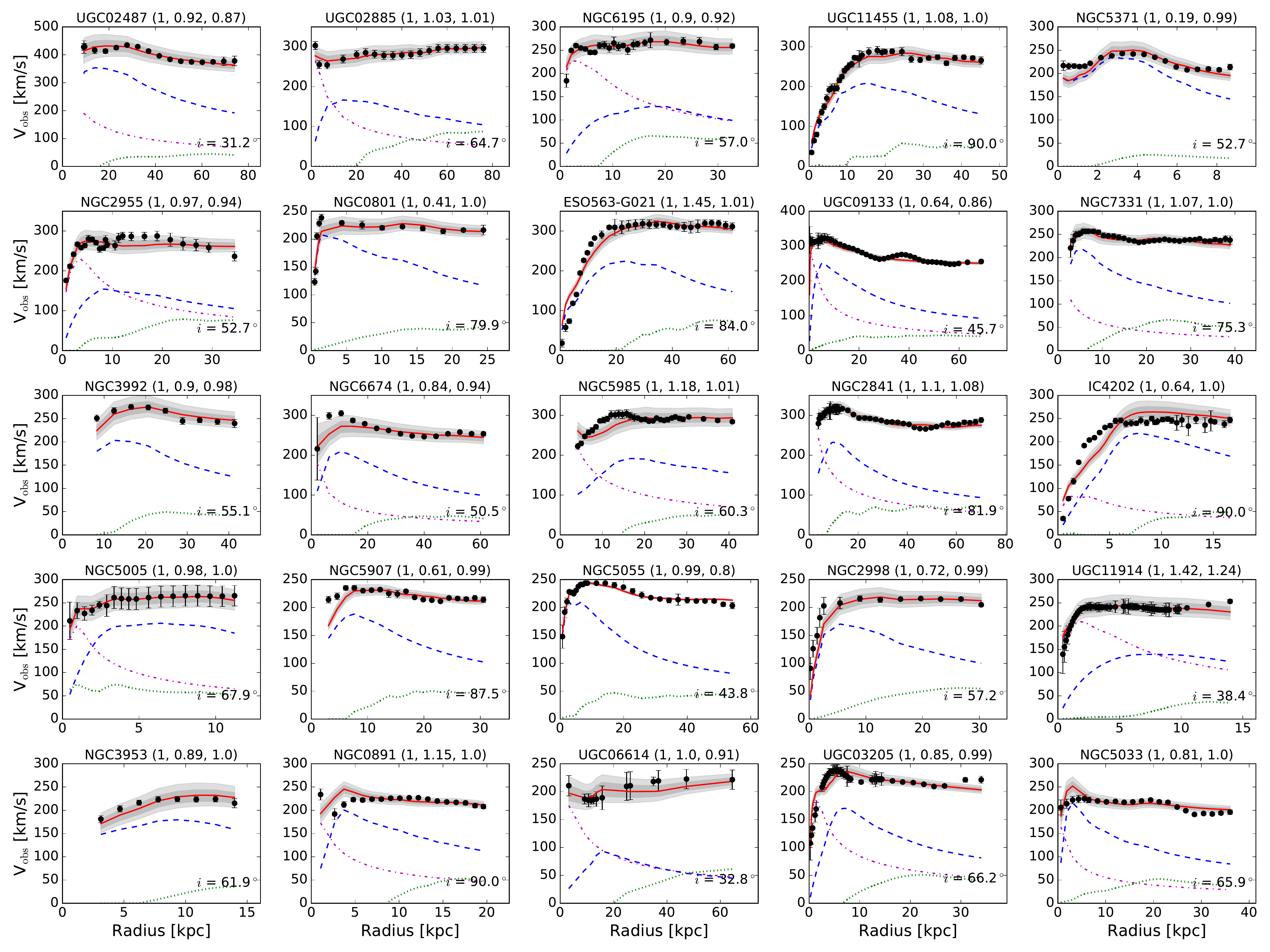}
\includegraphics[width=.9\columnwidth]{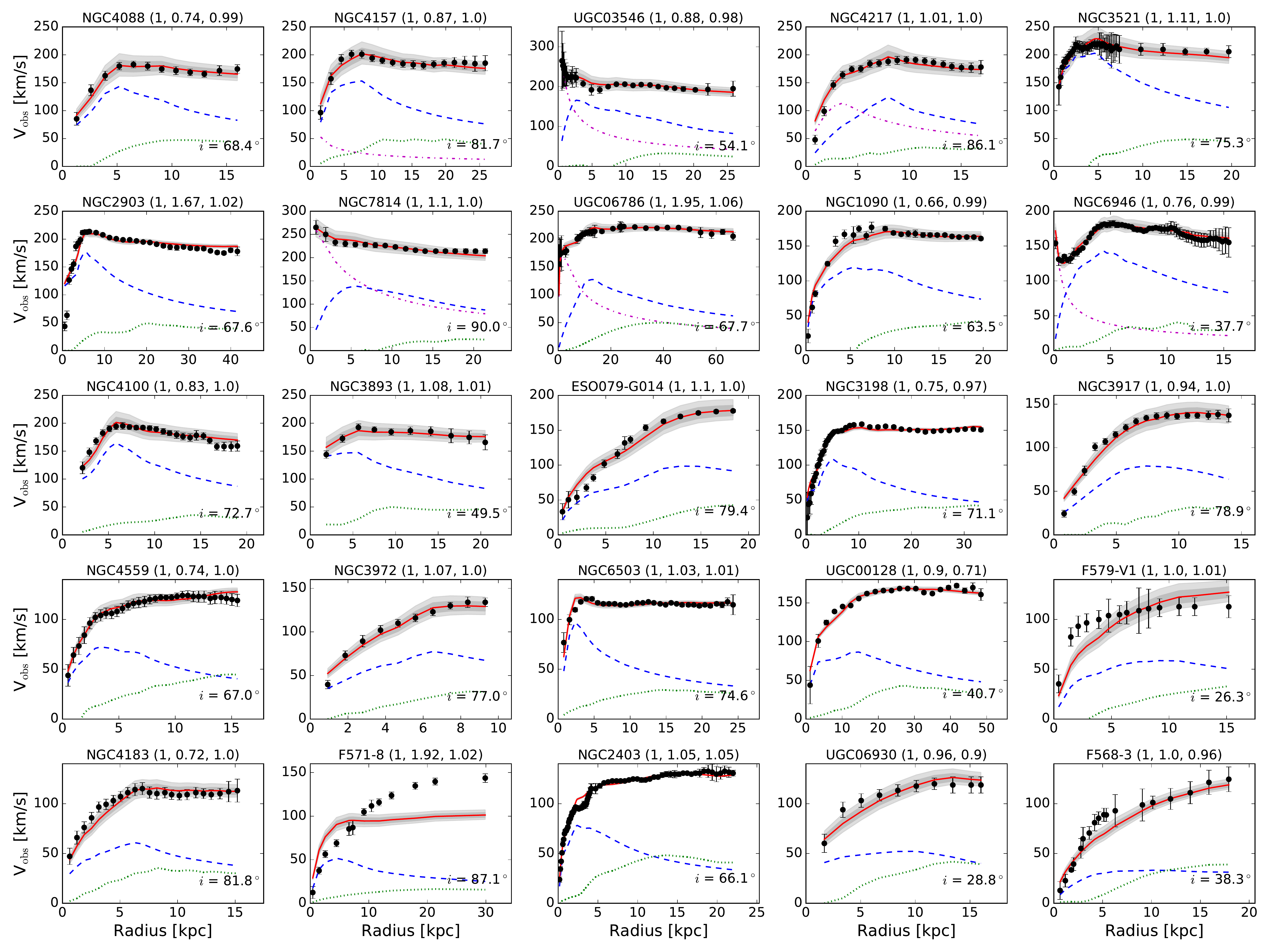}
\caption{MOND rotation curve fits of the 100 higher-quality galaxies from the SPARC sample as described in the text, shown as solid lines. The other curves show that Newtonian rotation curves calculated for the separate baryonic components: stars in the disc (dashed), and in the bulge (dashed-dotted), and cold gas (dotted).
Above each frame are given the galaxy name, and in parentheses: the quality grade, the distance adjustment, and the inclination adjustment that give the best fit (Li \& McGaugh, private communication).\label{sparc}}
\end{figure}

Two recent such studies are of particular note.
The first, by Li \& al., 2018, is notable because it included a large number (175) of disc galaxies, and because it employs a far-infrared light distribution, which is thought to be a good indicator of the stellar-mass distribution, better than the light at other wavelengths.
The results of this analysis is shown (in part) in
Fig. \ref{sparc}. Shown are the MOND fits for 100 disc galaxies from the SPARC sample.\footnote{These 100 galaxies are the ones with the higher-quality data; i.e., with $Q=1$ according to the authors' quality classification (the last galaxy has $Q=2$).} The data and analysis are described in Li \& al., 2018, and the Figures provided to me by Pengfei Li and Stacy McGaugh (private communication).

\begin{figure}[!ht]
\includegraphics[width=.9\columnwidth]{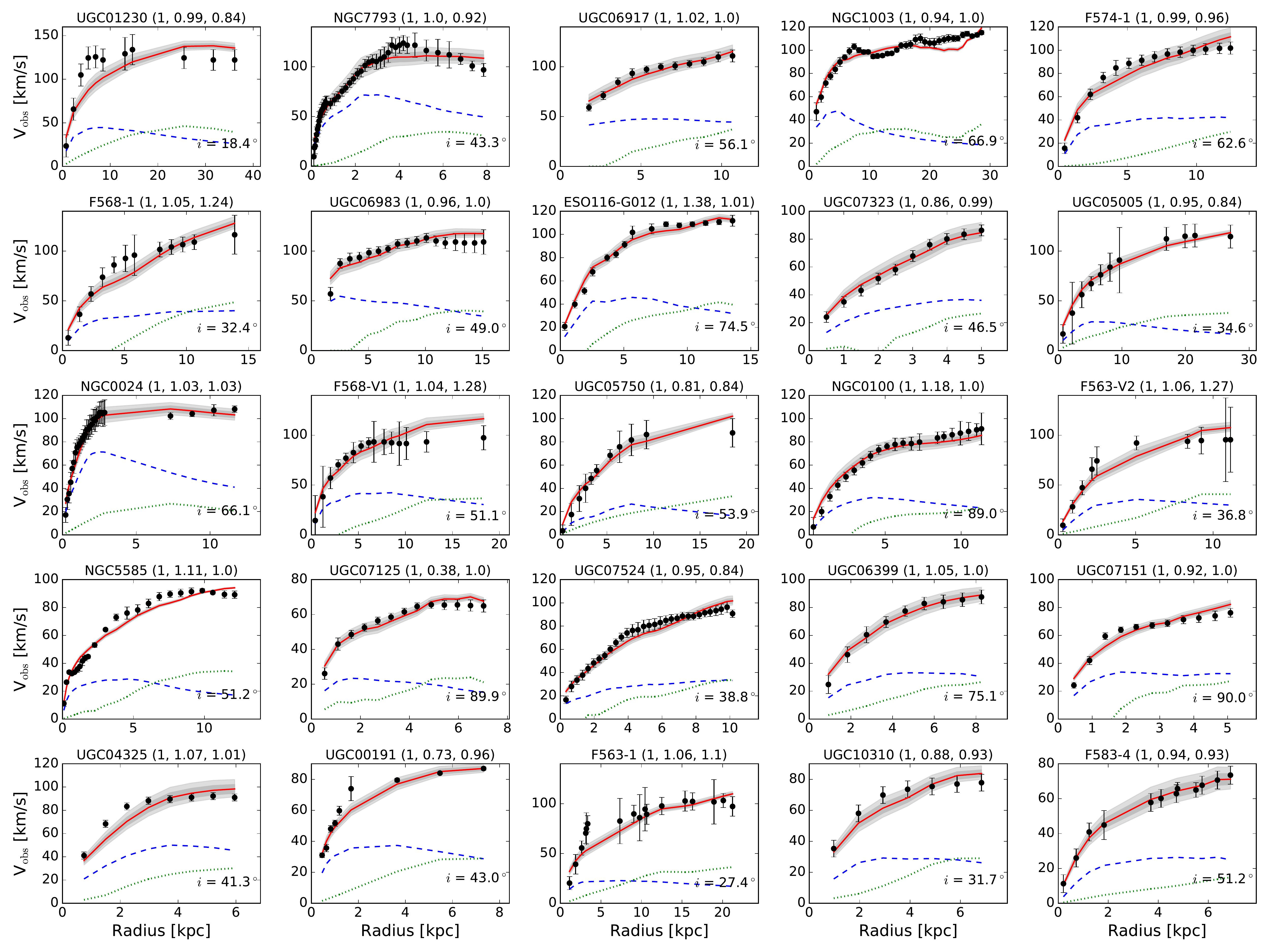}
\includegraphics[width=.9\columnwidth]{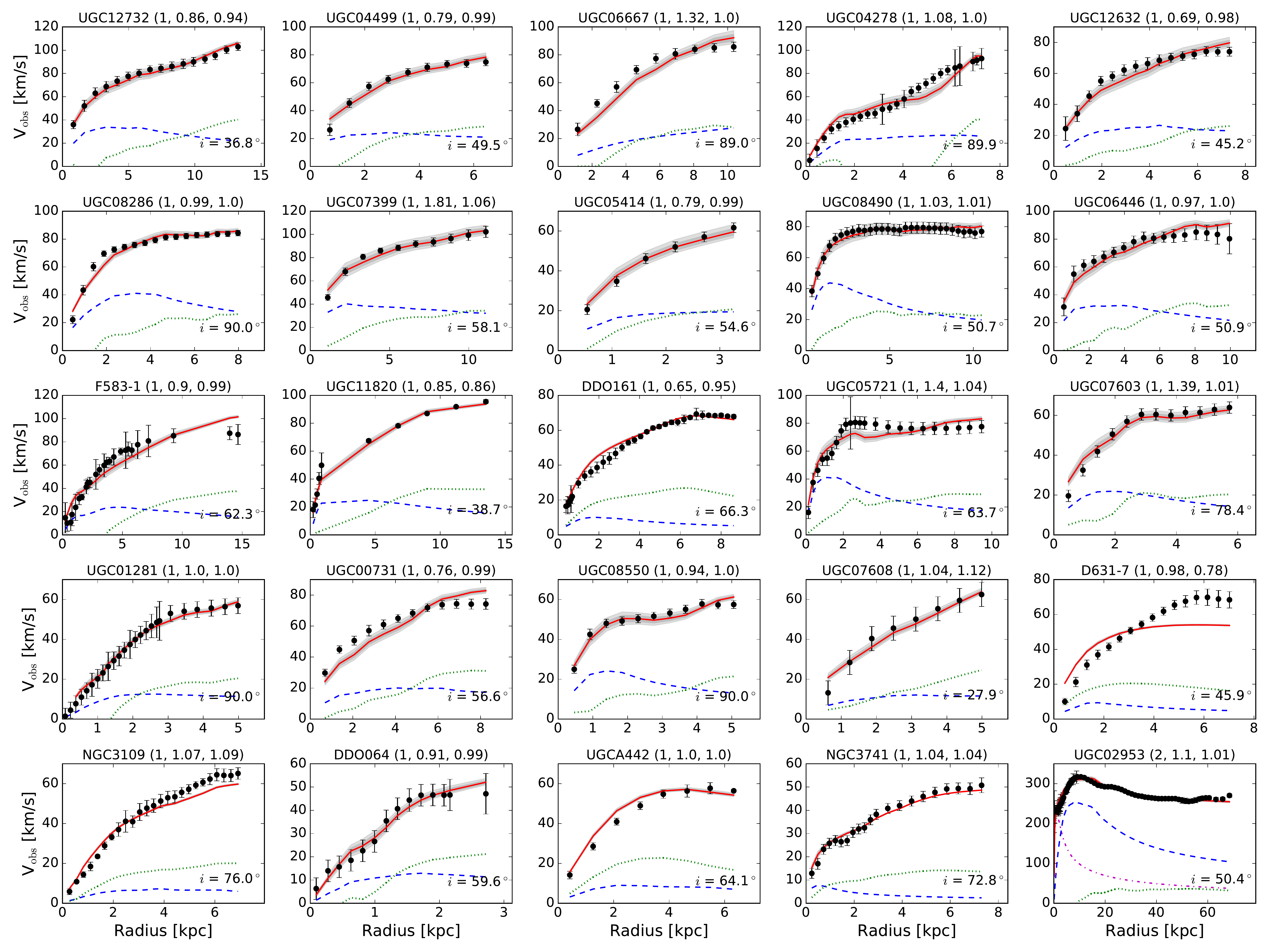}

Fig. \ref{sparc} (cont.)
\end{figure}

\begin{figure}[!ht]
\includegraphics[width=1.\columnwidth]{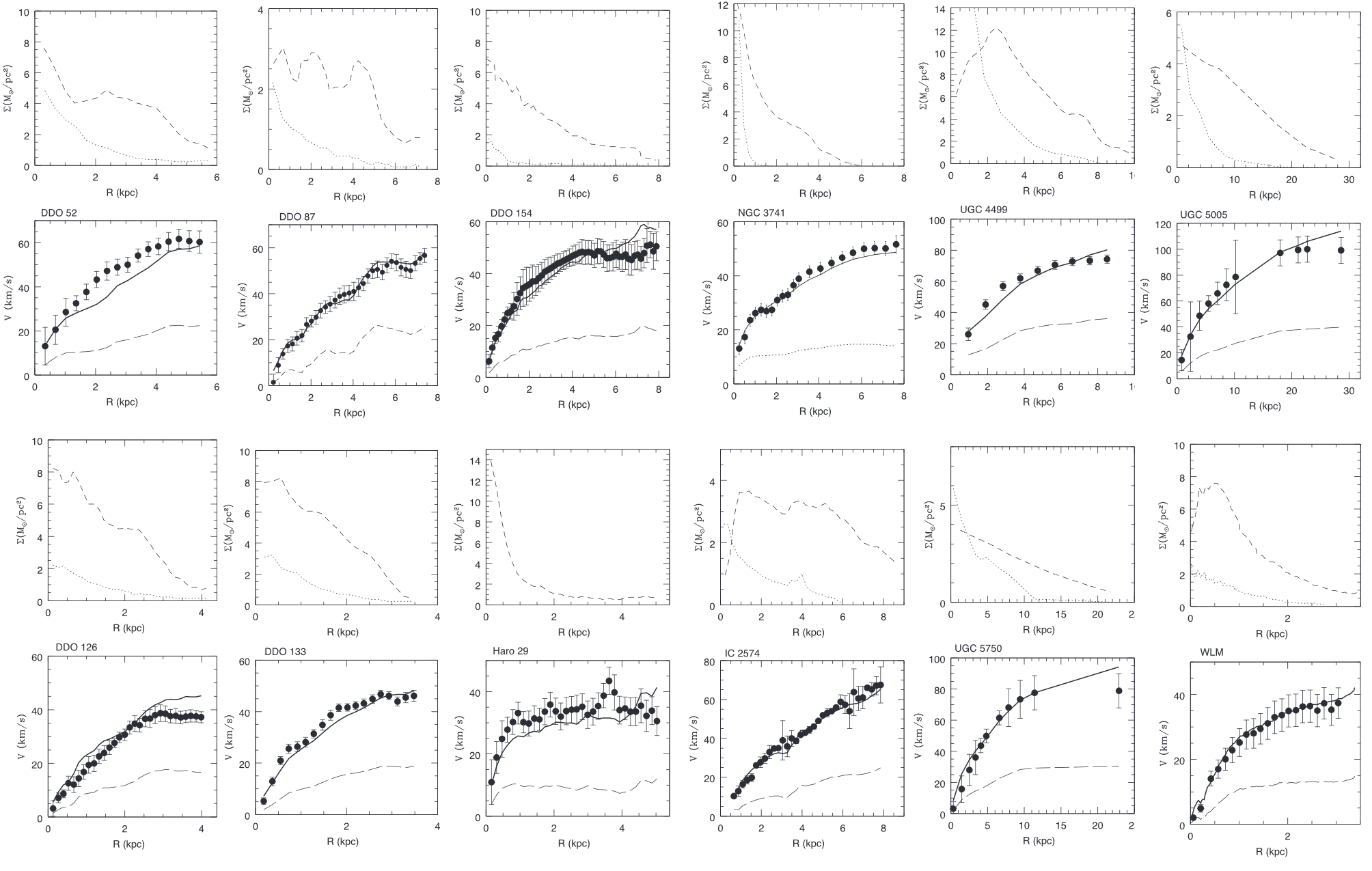}
\caption{MOND predictions of rotation curves of gas-rich, disc galaxies from Sanders, 2019. For each galaxy, the upper panel shows the surface densities of gas (dashed) and stars (dotted) as functions of radius. The lower panel shows the observed rotation curve (points), the Newtonian rotation curve of baryonic components (long dashed), and the predicted MOND rotation curve (continuous line). Because stars contribute very little to the baryonic mass (as is seen in the upper panel), the exact conversion of starlight to stellar mass is practically immaterial. To boot, the mass discrepancies are very large everywhere in these galaxies (as seen in the lower panels: the observed velocities are much higher than the baryonic ones). So we are fully in the DML, and the MOND predictions depend only very little on the exact form of the interpolating function.\label{sanders19}}
\end{figure}

The second recent study is by Sanders, 2019, whose results are shown in Fig. \ref{sanders19}. It uses a small, but select sample in two important regards: a. It involves only galaxies whose baryonic mass is dominated by gas. Since  stars make only a relatively small contribution to the baryonic mass the uncertainty introduced by converting starlight to stellar mass is largely obviated. b. The galaxies are each fully in the DML, making the MOND prediction independent of the exact form of interpolating function. Some of the galaxies in this sample where analyzed in other studies before, with similar success.

\subsubsection{Dwarf spheroidal satellites of the Milky Way and Andromeda--internal dynamics\label{dwarfs}}
Dwarf spheroidal satellites of the Milky Way and Andromeda exhibit large mass discrepancies. Within Newtonian dynamics they all require dominant contributions of DM to their self gravity. These dwarfs also all have internal accelerations well below $\az$. This latter fact was evident already at the time of the advent of MOND, based on their observed low surface densities. And so we find in Milgrom, 1983: `The best potential test of these suggestions I can think of involves the dwarf elliptical galaxies in the vicinity of the Milky Way', and in Milgrom, 1983a: `...we predict that when velocity dispersion data is available for the dwarfs, a large mass discrepancy will result when the conventional dynamics is used to determine the masses.'
\par
The most detailed MOND analyses of Milky-Way dwarf satellites, accounting for the full radial dependence of the line-of-sight velocity dispersions, was done in
Angus, 2008, and in Serra et al., 2010. Out of the eight dwarfs studied -- almost all of which conform well with the predictions of MOND -- two stand out as possibly being in tension with MOND: Carina and Draco. While the shape of the velocity distribution can be reproduced with MOND, these two galaxies still require too large mass-to-light ratios for stars alone.
\par
Read \& al., 2019 focused on these two rogues as a test of MOND, and also find tension with MOND for them. In particular, they make the point that these two galaxies have similar light density distributions but different dynamics. This, they say, is not expected in a theory like MOND, where the baryon distribution determines the dynamics.
\par
We have to remember, however, that strong assumptions always underlie dynamical analyses, for example, that the studied systems are in a quasi-static state of equilibrium. It may well be that some of the dwarfs are not in equilibrium, but in a perturbed dynamical state due to perhaps a past interaction with the Milky way, or a recent merger they underwent.
If such a breakdown of the assumptions is responsible for an apparent conflict with MOND in these two dwarfs, then this breakdown would have occurred in different ways in the two. So the fact that they show different dynamics is not meaningful in this scenario.
\par
There are also fainter satellites of the Milky Way that afford only global analyses.
They, however, tend to be in a perturbed, nonequilibrium state (e.g., subject to ongoing tidal disruption) in which case the standard dynamical analysis is not applicable.
A discussion of these fainter dwarfs in MOND can be found in McGaugh \& Wolf, 2010.
\par
\begin{figure}[!ht]
\begin{tabular}{ll}

\includegraphics[angle=-90,width=.95\columnwidth]{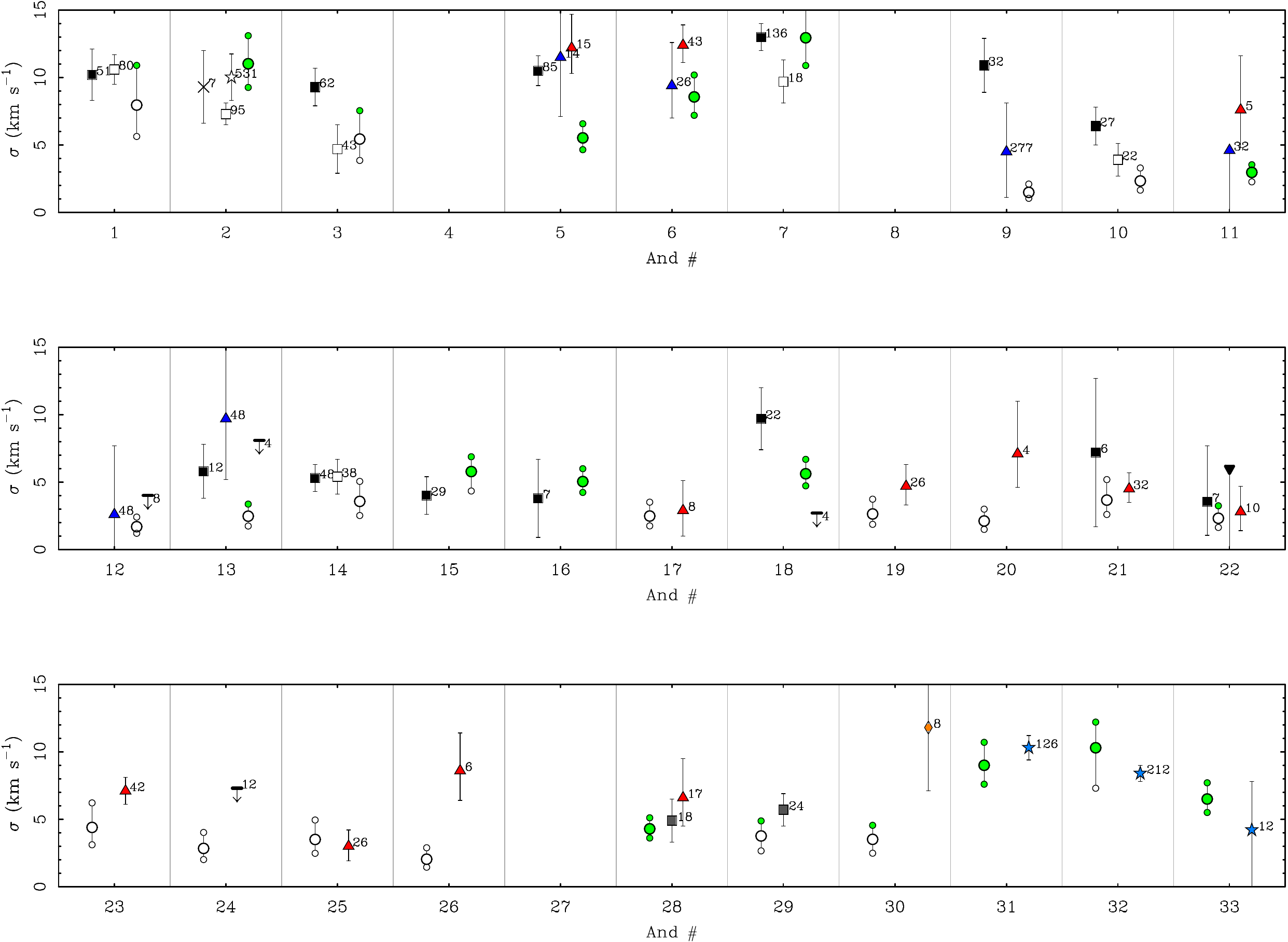}
\end{tabular}
\caption{MOND predictions of the line-of-sight velocity dispersions of 30 dwarf-spheroidal satellites of Andromeda, shown as circles with some estimate of the error margin (McGaugh \& Milgrom, 2013a, 2013b, and  Pawlowski \& McGaugh, 2014). Also shown within each numbered frame are actual measurements and upper limits (as squares and triangles).
The left-to-right order of the points for each dwarf is the actual temporal order in which the results were obtained. Numbers next to data points give the number of stars on which the measurement is based.\label{anddwarfs}}
\end{figure}

The dwarf-spheroidal satellites of our neighbor, the Andromeda galaxy, are typically about 10 times farther than the Milky Way satellites, and so have afforded only global analysis:
Only integrated line-of-sight velocity dispersions are available; so only total masses within the stellar body can be derived (estimated). The measurement of velocity dispersions, and the MOND predictions from the observed luminosities of these are shown in Fig. \ref{anddwarfs} (using the canonical value of $\az= 1.2\times 10^{-8}~\cmss$). For many of these, the MOND predictions were made before the measurements (as indicated by the order of the marks in the figure). Here and there we see apparent disagreements between the MOND predictions and measurements (especially in measurements based on a small number of stars, which are not reliable). But given the many sources of systematic errors this results bespeak success of the predictions.

\subsubsection{Elliptical galaxies}
Elliptical galaxies are not as amenable to dynamical analysis as are disc galaxies.
With few exceptions, one has to use test particles -- such as the stars, planetary nebulae, and globular clusters -- whose orbits are not known, unlike the nearly circular orbits in disc galaxies, and which do not probe the dynamics to large radii.
(See more details and further references in Milgrom, 2012.)
\par
Weak gravitational lensing, discussed above, does overcome these issues, but gives only statistical averages for many galaxies.
\par
The MDAR analysis of pressure-supported systems discussed above in Sec. \ref{masrpress}
includes also elliptical galaxies, but, again, treated statistically.
\par
An extensive analysis of the dynamics of individual elliptical galaxies in MOND, using their globular-cluster systems as test-particle probes, are discussed in B\'{i}lek \& al., 2019a, 2019b.
\par
Some cases are known of isolated elliptical galaxies shrouded by a halo of hot, x-ray-emitting gas. Analysis of the radial temperature and intensity distributions of the x-rays can be used to deduce the radial run of the gravitational acceleration. In such analysis one assumes that the x-ray-emitting gas is everywhere in hydrostatic equilibrium (pressure gradients balancing gravity).
\par
Even though such systems are rare, I mention them here because they are also rare examples of the gravitational field of individual elliptical galaxies being probed to large radii and low accelerations.
\par
\begin{figure}[!ht]
\begin{tabular}{ll}
\includegraphics[width=0.45\columnwidth]{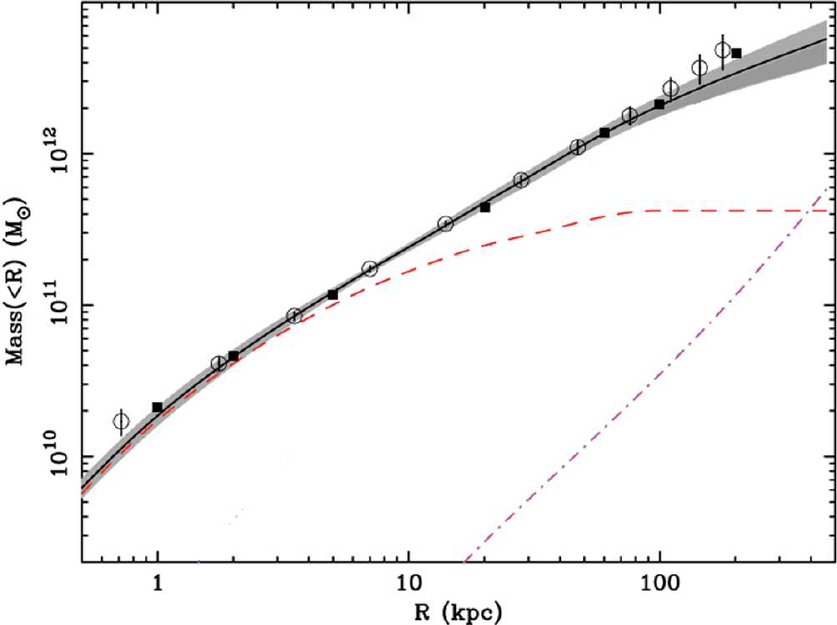}
\includegraphics[width=0.45\columnwidth]{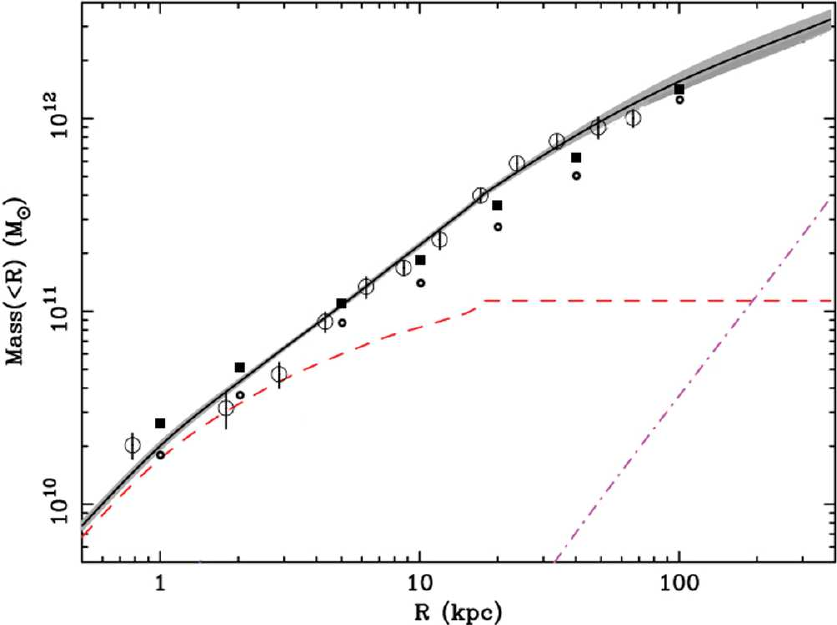}
\end{tabular}
\caption{Dynamical analysis of two isolated elliptical galaxies, using observations of their x-ray emitting, hot-gas envelopes. The different measures of mass enclosed within radius $r$ as a function of $r$: That of stars (dashed) and hot gas (dot dashed); the dynamical masses (grey band and large circles) deduced from observations and different analyses of the x-ray distribution from Humphrey \& al. (2011,2012); and the values predicted by MOND from the baryon distribution (squares and small rings). From Milgrom, 2012. \label{xray}}
\end{figure}

Figure \ref{xray} shows the MOND predictions of the gravitational acceleration (expressed as the mass needed to produce it in Newtonian dynamics -- the so called `dynamical mass') as a function of radius.  Also shown are the values expected from Newtonian analysis without DM, and those deduced from analysis of the distributions of x-ray emission. In both cases it is clear that large mass discrepancies appear beyond radii of about 10 kiloparsecs, increasing to a discrepancy of a factor 10-20. MOND reproduces these observations very well.
Buote \& Barth, 2019 have recently claimed that similar analysis of a third such case gives results in tension with MOND.

\subsubsection{Medium-richness galaxy groups\label{groups}}
Medium-richness galaxy groups are comprised typically of a few tens of galaxies, with the main baryonic mass in the form of stars and perhaps some cold gas. I have discussed their MOND dynamics in several papers, most recently and extensively in Milgrom, 2019, using data from Makarov \& Karachentsev, 2011.
\par
Each such group affords at present only global analysis.
And, as individual systems, they are not so amenable to dynamical analysis because of various uncertainties that beset the interpretation of the data.  However, as an ensemble, they provide a very important test of MOND and DM: First, they constitute a class of objects distinct from galaxies and galaxy clusters in how they form and evolve, with possibly different roles of baryons (and the putative DM) (see e.g., Oehm \& al., 2017); they thus enlarge the testing grounds of MOND in a significant way.
\par
Such groups are much larger in size than galaxies, but comparable in size to cores of galaxy clusters. The accelerations in them are typically much lower than within galaxies and clusters: down to $\sim 10^{-2}\az$, compared with  $\sim 10^{-1}\az$ that can be probed within galaxies. They thus afford testing MOND at much lower accelerations than is otherwise possible, and at radii as large as those probed in galaxy clusters.
This extension of the tests are important because it has been suggested that MOND as is now formulated may break down at very low accelerations or at large radii.
Groups dynamics as analyzed in Milgrom, 2019 and earlier analyses, are well-consistent with the predictions of MOND, thus negating both these claims.
\par
\begin{figure}[!ht]
\begin{tabular}{lll}
\includegraphics[width=1.\columnwidth]{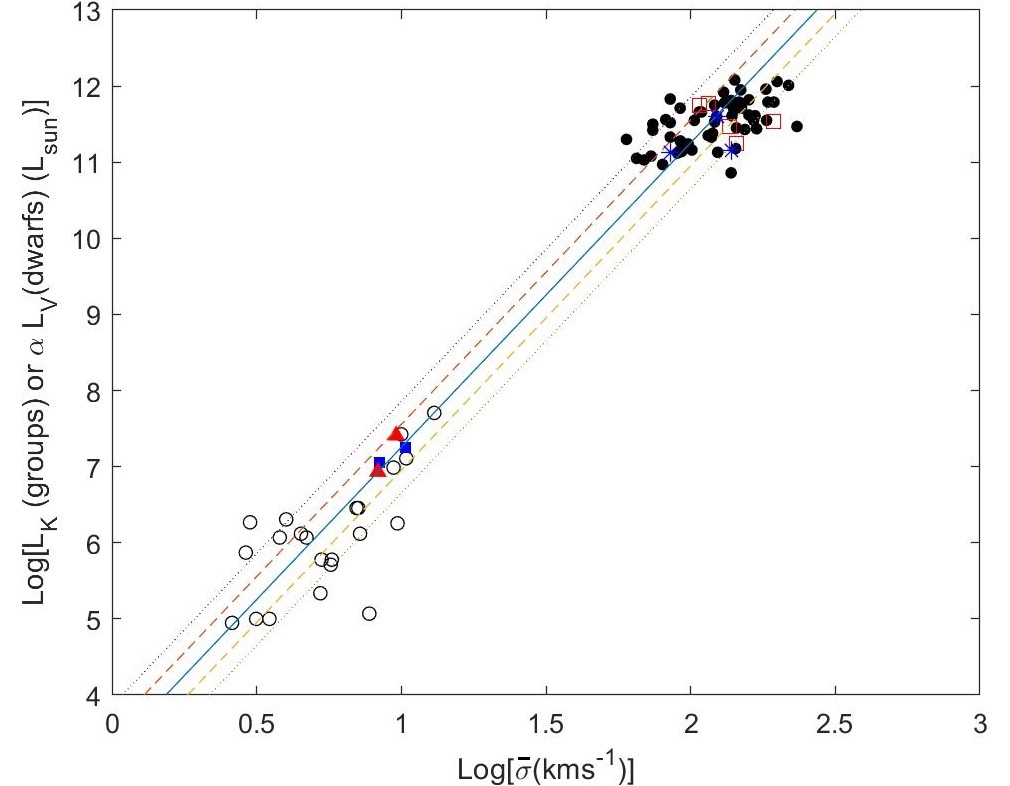}
\end{tabular}
\caption{The luminosity (in solar units) for the groups and dwarfs plotted vs. the line-of-sight velocity dispersion, $\bar\s$. Groups (data from Makarov \& Karachentsev, 2011): filled circles, open squares, and asterisks.  Dwarfs: open circles, filled squares, and filled triangles.  The lines show the predictions of law (iv) [eq. (\ref{msigma})] with $q=4/9$ -- which is the value predicted by modified-gravity MOND theories -- for several reasonable values of the mass-to-light ratios of the groups and dwarfs. Details and references for the data in Milgrom, 2019.
$\s$ in eq. (\ref{msigma}) is the full 3-dimensional velocity dispersion and is statistically related to $\bar\s$ as $\s^2=3\bar\s^2$.\label{msigmagroups}}
\end{figure}
\begin{figure}[!ht]
\begin{tabular}{lr}
\includegraphics[width=1.\columnwidth]{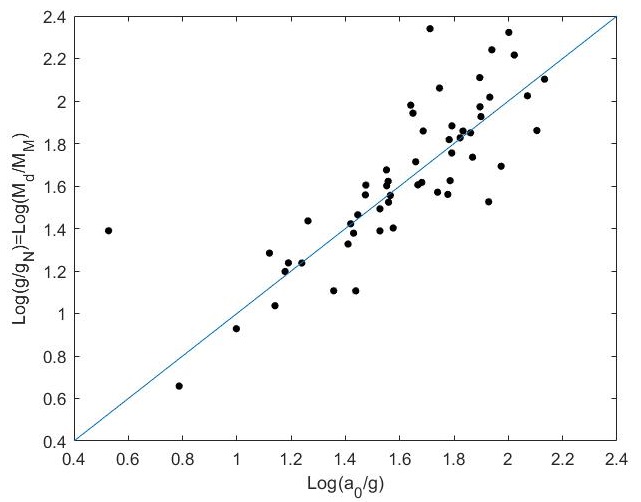}
\end{tabular}
\caption{The MDAR for galaxy groups: The mass discrepancy -- i.e., the ratio $g/\gN$, of the measured acceleration, $g$, to the Newtonian, baryonic one, $\gN$, plotted vs. $\az/g$. The equality line -- which is the MOND prediction for such very low accelerations -- is also shown. \label{groupsmdar}}
\end{figure}

The main results of Milgrom, 2019, comparing the data with the predictions of MOND for the canonical value of $\az$, are shown here in Figs. \ref{msigmagroups} and \ref{groupsmdar}.
Figure \ref{msigmagroups} shows results for both dwarf satellites of Andromeda (discussed in Sec. \ref{dwarfs} above), and the groups, because it is the same MOND prediction that is used in the analysis of both, and because the comparison is revealing. Each system is plotted in the $\bar\s-L$ plane ($L$ is the luminosity of the system and $\bar\s$ the measured line-of-sight velocity dispersion). Also shown are the predictions of MOND's law (iv), [eq.(\ref{msigma})] for several reasonable values of the mass-to-light ratio. [Law (iv) predicts a relation between $\s$ and the baryonic mass; we need this conversion factor to get the latter from the observed luminosity.]
\par
The main lesson is that galaxy groups and dwarfs follow the same MOND relation.
This is particulary significant, given that the groups are hundreds of times larger in size, and millions of times more massive than the dwarfs, and that the two types of systems must have had very different processes governing their formation and evolution.
\par
Figure \ref{groupsmdar} is the MDAR for galaxy groups. It plots the same quantities as in Figs. \ref{mdar1} and \ref{scarpa}, for the group sample.
The line in Fig. \ref{groupsmdar} is the continuation of the predicted, low-acceleration MOND asymptotes shown in Fig. \ref{mdar1}.
Thus, the points in Fig. \ref{groupsmdar} fall on the MOND prediction in Fig. \ref{scarpa} for abscise values there between $10^{-10}\cmss$ and  $10^{-12}\cmss$. This plot is also
an extension of Fig. \ref{mdar1} by two orders of magnitude down to Newtonian accelerations of $\approx 10^{-14}~\mss$. It shows that the MOND predictions for groups hold down to such low accelerations.

\subsubsection{Galaxy clusters\label{clusters}}
MOND does not fully account for the mass discrepancies in galaxy cluster and rich galaxy groups that contain large quantities of x-ray emitting, hot gas. This was first pointed out in relation to the Coma cluster in The \& White, 1988, and then shown to apply to other clusters and rich groups in several publications.
\par
\begin{figure}[!ht]
\includegraphics[width=1.0\columnwidth]{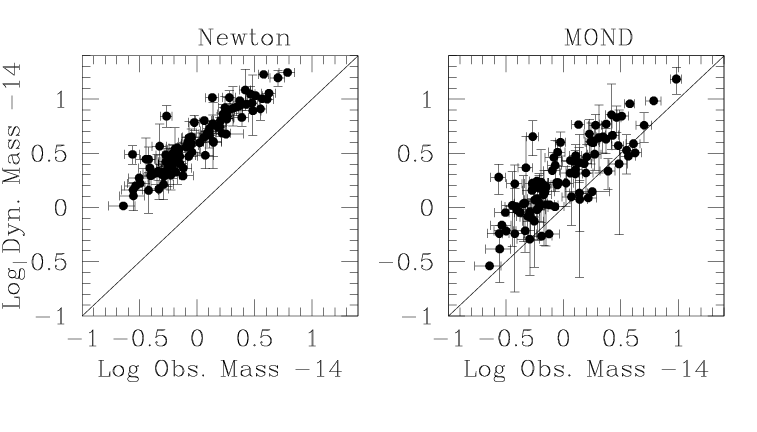}
\caption{Galaxy clusters: The dynamical mass (within some fiducial radius of about 1 megaparsec) -- i.e., the mass needed to balance the observed motions -- plotted against the observed baryonic mass for many galaxy clusters (each represented by a data point); taken from Sanders, 1999. On the left are plotted the dynamical masses required by Newtonian dynamics, which are seen to be about five times larger than the observed masses. On the right are the masses required by MOND, which are much reduced, but still about a factor of two too large.}\label{clusterssanders}
\end{figure}

The global mass discrepancy in clusters in Newtonian dynamics is of a factor of about 5-10. MOND does reduce it greatly but not completely: Even with MOND there is a remaining discrepancy of about a factor of 1.5-2. Figure \ref{clusterssanders}, from Sanders, 1999, shows these two discrepancies by plotting the required dynamical mass in Newtonian dynamics and in MOND vs. the presently observed baryonic mass.
Similar results were reached more recently by Ettori \& al., 2019.
\par
Unlike apparent tensions with MOND that appear occasionally in individual systems and that can generally be attributed to suspected or unknown systematics, here the discrepancy is quite ubiquitous and requires more serious consideration.
Possible solutions to this conundrum have been brought up, including some modifications of the present formulation of MOND (e.g., introducing dependence of $\az$ on environment).
\par
The minimalist explanation -- discussed in detail in Milgrom, 2008 -- is that galaxy clusters and rich groups that bear large quantities of baryonic matter in the form of hot, x-ray emitting gas, bear also some yet-undetected baryonic component (such as dead stars or cold gas clouds). To bridge the remaining discrepancy in MOND, this component has to be roughly as massive as the observed gas, and be distributed roughly like the galaxies in the cluster (which are more centrally concentrated than the gas). Such additional baryons will only constitute about five percent of the baryonic budget in the Universe.
\par
Furthermore, with these attributes, this baryonic component would also account for the observations of the `Bullet Cluster', which has been bruited as strong evidence for DM  (Clowe \& al., 2006).

\subsection{Cosmology \label{cosmology}}
Cosmology deals with the Universe at large as a dynamical system. It strives to account for the geometry of the Universe's spacetime (e.g., how space expands as a function of time). Cosmology also deals with the behavior of the material content of the Universe, and how it affects and is being affected by the expansion, including the initial stages of departure from homogeneity, and the formation of structure. Gravitational aspects of this system are conventionally treated within GR.
\par
As stated above, this paradigm clearly conflicts with observations unless we invoke two disparate dark components, DM and `dark energy'. Their supposed properties differ greatly. For example, DM can only slow down the expansion of space. So the fact that the expansion was observed to accelerate has called upon `dark energy' that causes such accelerated expansion. A `cosmological constant', is a straightforward addition to GR that with the right sign can serve the purpose of `dark energy'.
\par
Because MOND has already introduced an acceleration constant, which observations tell us is of the order of the observed cosmic acceleration [eq. (\ref{coinc})], MOND theories can naturally account for `dark energy'. For example such a constant appears almost automatically in the `Einstein aether' formulations of MOND (Zlosnik, Ferreira, \& Starkman 2007), or in the bimetric formulation of MOND (Milgrom, 2009b).
\par
As regards cosmological DM, there are elements in MOND that could account for some roles of such DM (for example, in accelerating collapse of primordial matter fluctuations to form galactic objects).
\par
According to C. Skordis,\footnote{Private communication, and a talk presented at `BonnGravity2019 – The functioning of galaxies: challenges for Newtonian and Milgromian dynamics', held in Bonn in September 2019.} a subclass of the TeVeS extensions by Skordis \& Z{\l}o\'{s}nik, 2019, discussed in Sec. \ref{theory}, accounts for the observed cosmic-microwave-background-fluctuation spectrum, for the observed expansion history of the Universe, and for the matter power spectrum, without dark matter (a cosmological constant is still introduced by hand, as in $\Lambda$CDM).
\par
There are also MOND-DM hybrids that invoke DM for cosmology and large-scale structure, while striving to reproduce MOND phenomenology in galactic systems (see Sec. \ref{hybrids} below).
\par
Some aspects of cosmology and structure formation have been studied in existing relativistic MOND theories (e.g. Skordis, 2009; Clifton \& Zlosnik, 2010; Milgrom, 2009b, 2010; Clifton \& al., 2012; Z{\l}o\'{s}nik \& Skordis, 2017, and the above-mentioned conference paper by Skordis).
\par
In the context of GR, one first developed the theory of relativistic gravitation for local systems, and then it proved possible to apply this theory to the Universe at large, as a special case. This will not necessarily turn out to be the case in MOND.
In my pinion, relation (\ref{coinc}) between the value of $\az$ that appears in local, galactic dynamics and cosmological accelerations, may be telling us that we will need to understand local MOND and cosmology as inseparable aspects of the same FUNDAMOND theory underlying our present notion of MOND. If so, it may be futile (but not deprecated or discouraged) to try and understand cosmology within some local theories of MOND.
\par
A still heuristic, but revealing, model based on a picture whereby our observed four-dimensional space time is a manifold embedded in a higher dimensional space time -- so called `membrane world picture' -- is described in Ref.  \cite{milgrom19a}. It accounts naturally for the `dark-energy' acceleration and its relation to $\az$ as in eq. (\ref{coinc}). It also gives a possible solution to the so-called `old cosmological-constant problem', which refers to the fact that the very large energy density of the quantum vacuum does not show up as a very large cosmological constant.

\section{Dark matter?\label{DM}}
{\it ``Whatever difficulties we may have in forming a consistent idea of the constitution of the aether, there can be no doubt that the interplanetary and interstellar spaces are not empty, but are occupied by a material substance or body, which is certainly the largest, and probably the most uniform body of which we have any knowledge.'' }(J. C.
Maxwell, in `Ether', Encyclopedia Britannica, Ninth Edition).

{\it ``Does our ether actually exist? We know the origin of our belief in the ether. If light takes several years to reach us from a distant star, it is no longer on the star, nor is it on the earth. It must be somewhere, and supported, so to speak, by some material agency.''} (H. Poincar\'e, in `Science and Hypothesis').
\par
Indeed, the arguments for the existence of the aether had been no less compelling and cogent than those for the existence of DM today. The aether's existence was a generally accepted fact; yet, like epicycles, the phlogiston, and the caloric, it proved to be only a figment.
\par
Dark matter and its performance has been extensively reviewed and congratulated for its performance (see, e.g. the references given in the Introduction).
Here I emphasize some of the weaker aspects of this paradigm.

\subsection{Some dark matter worries \label{worries}}
{\it The overall impression one obtains, given the best data we have, is matter to be arranged as not expected in the dark-matter based standard model of cosmology...This evidence suggests strongly that dynamically relevant dark matter does not exist and therefore cosmology remains largely not understood theoretically.} (Kroupa, 2016)
\par
{\it `There is a growing sense of `crisis' in the dark-matter particle community, which arises from the absence of evidence for the most popular candidates for dark-matter particles such as weakly interacting massive particles, axions and sterile neutrinos despite the enormous effort that has gone into searching for these particles.} (Bertone \& Tait, 2018)
\par
If one does not presuppose anything about what DM is, about what its interactions with baryons and with itself are (besides gravity), and about how it got to where it is, then it is hard to rule it out (barring some yet unsettled possibilities discussed in Sec. \ref{discriminants}).
\par
For many years, most of the DM community has committed itself to a class of candidates designated `cold DM' (CDM), or `Lambda CDM' ($\Lambda$CDM) for the paradigm that includes dark energy.
However, in recent years, and in light of the difficulties facing DM in general, but \LCDM in particular, this committal is breaking down (see Sec. \ref{balkan}.)
\par
There are several nagging worries that beset the DM picture. None of them has proven fatal; but together they certainly give one pause. To my mind, the main ones are as follows:
\par
1. No known form of matter can be the DM.

The standard model of particle physics accounts very well for all forms of matter known to exists. With few exceptions (such as non-zero masses of the neutrinos) it also accounts for the properties of this matter.
There is, however, no room in the standard model for anything that could constitute the DM, which must then be made of a yet unknown substance.
All `standard-matter' candidates, such as dead stars and massive neutrinos, that have been considered over the years have all been ruled out.
Because of this and other shortcomings, there are hypothetical suggestions to extend the standard model, some better, some less motivated. Some of these extensions involve new types of particles, some of which might be DM constituents. But there is not  a shred of evidence that such particles do exist.
\par
2.
Many experiments attempting to detect DM (directly and indirectly) have come up empty.

Since our galaxy is thought to be enshrouded in DM,
ingenious ways have been devised to detect directly the DM particles that swarm by the Earth.
Many such experiments have come and gone, and quite a few are still afoot.
\par
In addition there have been many attempts to detect secondary effects of DM (via so-called `indirect detection') such as the electromagnetic radiation given off by the decay or annihilation of DM.
\par
3.
Contrary to hopes (and expressed certitude by some), no candidate DM particle or a hint of any `physics beyond the standard model' that could tell us indirectly that such DM exists, have come up (e.g., in the Large Hadron Collider).
\par
4.
At face value, at least, many observations of galaxy structure and dynamics conflict with natural predictions of \LCDM (e.g., Peebles \& Nusser, 2010; Kroupa 2015, 2016; Wu \& Kroupa, 2015;
Bullock \& Boylan-Kolchin, 2017;  Del Popolo \& Le Delliou, 2017; Pawlowski, 2018).
These conflicts go under names such as `the cusp-core problem', `the missing satellite problem', `the angular momentum catastrophe', `the  problem  of  satellites  planes', etc.
Attempts to explain these conflicts away are discussed in some of the above references.
\par
5.
Invoking dark matter alone is not enough to account for the observations, which require another ad-hoc addition to standard dynamics in the form of `dark energy'.
\par
6.
There are unexplained `coincidences' relating to the properties of the DM and dark energy. For example, the amount of matter in the universe required in GR to explain the observed cosmology is approximately six times the amount of baryons we think exist. So the amount of DM needs to be about five times that of baryons.
\par
However, the quantities of DM and of baryons in the Universe are thought to have been determined and fixed at different cosmic epochs in the early Universe, and by unrelated physical processes. Why then should their quantities be so near in value? This might be a mere coincidence. But it may also be telling us that DM does not have life and existence of its own, that it is an artefact of some unrecognized effects of baryons -- exactly as MOND suggests. For example, in the GR equations that describe the expansion history of the Universe, matter density appears as $G\r(matter)$ and all we learn from observations of the expansion is that this quantity is too small by a factor of about $2\pi$ if we take the baryon density, $\r(baryons)$ for $\r(matter)$.\footnote{Both densities decrease with the expansion, but their ratio remains essentially the same.} The mainstream resolution is to invoke DM so that $\r(matter)\approx 2\pi\r(baryons)$. But, in the spirit of MOND it could also be that the effect of baryons is $2\pi$ times stronger than in GR (or that the relevant value of $G$ is $2\pi$ times larger in cosmology than the value in the solar system).
\par
Another well-known coincidence is that the total required matter density {\it today} is of the same order as that of the dark energy (in a ratio of about one to three).
\par
MOND has revealed another coincidence, expressed in eq. (\ref{coinc}), and discussed in detail above: $\az$, which appears
in so many aspects of galaxy dynamics, is related to the cosmic acceleration parameters.
\par
7.
Even where \LCDM performs well, i.e., in the area of cosmology and structure formation, not all is roses. For example, recent observations challenge the \LCDM structure-formation scenario, showing that various objects -- such as massive galaxies, quasars, and clusters of galaxies -- were already formed at much earlier cosmic times than \LCDM tells us they should have (e.g., Franck \& McGaugh, 2016; Steinhardt \& al., 2016; Hashimoto \& al., 2018; Timlin, \& al., 2018; and Wang \& al., 2019). This is discussed in more detail in McGaugh, 2018, with comparison with the expectations from MOND.
\par
8.
Present-day galaxies are the end results of haphazard, cataclysmic, and unknowable histories, involving collapse, collisions and mergers with other galaxies, star formation, explosion of stars, energetic activity in the galactic center, expulsion of most of the baryons from the galaxy, etc., etc.
But, the purported DM would have been affected very differently from baryons by all these processes.\footnote{For example, unlike baryons, DM is supposed to be neutral and not be affected by electromagnetic fields; it is very weakly interacting with baryons (and with itself in most versions), so it should be very ineffective in dissipating energy, etc. This results, for example, in baryons ending up in small rotating discs in disc galaxies, while the DM purportedly ends up in a much more extended spheroidal halo. We should also expect other clear manifestations of lack of correlations between the two components, much unlike what is observed.} Yet, as predicted by MOND, and as observed and discussed above, baryons alone are seen to determine the full dynamics in galaxies, even those aspects that would be governed predominantly by the purported DM. If the dynamics are described in terms of DM, one may express the observations as very tight correlations between the properties of baryons and DM, totally unexpected in the DM picture. See more on this in Sec. \ref{prediction}.
\par
To me, this observation is the most cogent argument against DM and for a unifying, modified-dynamic theory such as MOND.

\subsubsection{Balkanization of the DM paradigm\label{balkan}}
{\it `... in the early 1770’s, there were almost as many
versions of the phlogiston theory as there were pneumatic chemists. That proliferation of versions of a theory is a very usual symptom of crisis. In his preface, Copernicus complained of it as well.'} (Kuhn, 1962; in `The structure of scientific revolutions'.)
\par
Faced with the above conundrums -- especially with the so called `small-scale' (viz. galaxy scale) problems, and the failure of DM to show up in extensive searches -- the DM community is responding by putting on the table and exploring a large variety of DM candidates. Some of these were proposed to avoid some of the small-scale problems, some others were proposed in order to account for the continuing non-detection of DM, and some were proposed merely for the heck of it.
Because some of these, singly, do not cover all issues, cocktails of DM or whole `dark sectors' are also being discussed. Literally, tens of such DM options are now being discussed with varying degree of seriousness.
\par
These candidates can be classified in different ways: According to their material consistence -- be it fluid like (`Chaplygin gas', `superfluid DM', etc.), or made of macroscopic bodies (MACHOS, primordial black holes, `Macros'), or of microscopic particles.
Or they can be classified according to the strength of their interactions between themselves, or with baryons. DM particles can be, according to their mass, `hot' or `cold', or if necessary `warm'.
Some candidates can be described adequately as made of classical (non-quantum) constituents, while some are described as a macroscopic, globally-coherent quantum state (`fuzzy' DM). Candidates can also be classified according to how they emerge in various attempts to go beyond the standard model of particle physics. Thus we have `Axions', `Axinos', `Gravitinos', `photinos', or other `inos' from supersymmetric extensions. We also have `sterile neutrinos', `dark photons', `Kalutza-Klein DM', `Little Higgs DM', etc.
For partial reviews of such candidates see e.g., Feng 2010 (e.g., his Table 1) and Bertone \& Tait 2018 (e.g., their Fig. 1).
\par
This situation is in a sense the opposite of the `convergence' discussed in Sec. \ref{lawsanalog}.
As the latter is a good sign of health, proliferation to many sub-paradigms is a sign of weakness, as reiterated by Kuhn in the quote above.
The complaint that Kuhn alludes to regarding Copernicus's dismay with the proliferation of geocentric algorithms is another example:
{\it I was impelled to consider a different system of deducing the motions of the universe’s spheres for no other reason than the realization that astronomers do not agree among themselves in their investigations of this subject...they do not use the same principles,  assumptions, and explanations of the apparent revolutions and motions. For while some employ only..., others utilize..., and yet they do not quite reach their goal. } (Copernicus, preface to `De Revolutionibus', translated by E. Rosen.)
\par
Yet another example of a mainstream paradigm showing signs of falling apart through its balkanization is that of the nineteenth-century aether. This happened because it was impossible to find a simple aether model that could account for the accumulating facts. For example, the absence of aether drift in light abberation and in the Michelson-Morley experiment,  which required fluidity, and light polarization, which required a solid aether, etc. So, for instance, Stokes suggested an aether that behaves as a fluid on large scales, but as a solid on small scales -- of the order of light wavelength -- to be able to carry transverse waves.
As described in Pais, 1982:
{\it `However, there still were many nineteenth century candidates for this one aether... There were the aethers of Fresnel, Cauchy, Stokes, Neumann, MacCullagh, Kelvin, Planck, and probably others, distinguished by such properties as degree of homogeneity and compressibility, and the extent to which the earth dragged the aether along.}
\par
Those familiar with the many faces of the DM paradigm today will find close similarities with the state of the aether at the end of the nineteenth century.
\par
Why should we perceive Balkanization as a sign of weakness in a theory or a research program, and even a sign of a crisis looming?
In the first place, Balkanization is almost invariably driven by what Lakatos (1971) deprecated in a theory, and termed `degenerative problemshift'.\footnote{Merritt (2016) discusses the MOND vs. DM issue in light of these ideas by Lakatos.} Namely, the continual change of the premises of the theory in response to conflicting data. This might not be decisively fatal for a theory, but it is arguably a bad omen for its fate. A flowing stream changes its course where is encounters rocks. A good theory should, by and large, play the role of the rocks, directing and predicting the motion of the flow that is the observations, experiments, and data. It is a bad theory that is a stream that changes its course, and forks into rivulets, whenever it encounters obstacles.
\par
In addition, the multiplicity of the DM paradigm is in large part due to the fact that no single version of it, flexible as is, can avoid all the obstacles.
\subsection{MOND vs. DM: Predictions vs. after-the-fact explanations \label{prediction}}
{\it Given the obvious organization of the data, the many predictive successes of MOND, and the inability of \LCDM in many cases to make comparable predictions, it seems quite unreasonable to believe in the existence of invisible mass from beyond the Standard Model of particle physics that pervades the universe with nary a signal in the laboratory.} (McGaugh, 2015)
\par
As we saw, MOND has predicted many laws and regularities that have been vindicated by observations.
{\it None of these predictions had been made in the DM paradigm}.
On the contrary, as alluded to in point 4 in Sec. \ref{worries}, some natural predictions of \LCDM that concern the pure DM halos of galaxies conflict with observations, and now require elaborate, after-the-fact explanations of why perhaps things are not as the straightforward \LCDM predictions would have them.
\par
Even more significantly, and beyond the predictions of general laws,
MOND has made successful predictions of the dynamics of {\it individual} systems.\footnote{Predictions of laws would appear to include the predictions for individual systems. This would be true if the predicted laws are infinitely accurate. When DM attempts to make predictions of regularities, they are only correlations with large intrinsic scatter (added to the observational scatter due to measurement and systematic errors) but with individual predictions possibly significantly off.}
But DM is totally incapable of making predictions for individual systems without knowing the unknowable details of their history.
\par
These facts reflect a deeper, matter-of-principle difference between how MOND and DM account for the mass anomalies.
In MOND, all those laws and regularities of galaxy dynamics discussed in Sec. \ref{laws} are inevitable attributes of galaxies that have to be satisfied irrespective of how these galaxies formed and evolved. They are like Kepler's laws of planetary dynamics that follow from Newtonian dynamics and have to hold irrespective of how the planetary system formed.
In the DM paradigm this is not so: such regularities would have to depend crucially on the history of the individual galaxy under study.
\par
As regards the regularities MOND predicts, one finds in the literature claims that some of the correlations predicted by MOND can be reproduced in numerical simulations of galaxy formation and evolution (though they have never been predicted beforehand).
\par
These claims are, however, quite specious.
\par
a. These simulations are by no means ab initio calculations. They involve many free parameters and processes -- as alluded to in point 8 in Sec. \ref{worries} -- which are included in the simulations and adjusted by hand, with much freedom to affect the final products.
\par
b. Indeed, none of these results count as predictions. The simulations have been adjusted and readjusted over many years to cause the resulting galaxies to resemble as much as possible observed galaxies. And now we are told to behold and be impressed with how simulated galaxies resemble observed ones.
\par
c. Even so, different simulations do not agree with each other on the exact forms of these `resulting' correlations, and, to boot, they still are in conflict with observation (e.g., Wu \& Kroupa, 2015 vs. Tenneti \& al., 2018 vs. Keller \& Wadsley, 2017). For example,  Wu \& Kroupa, 2015, discuss the incompatibility of the results of various DM simulations with the observed MDAR. See also Milgrom, 2016b, who discusses various issues of principle; Tenneti \& al., 2018, who find that their simulations result in a MDAR unlike the observed one in that it has no break as shown in Fig. \ref{mdar1}; instead, it has a single (wrong) slope, and thus shows no sign of the critical acceleration $\az$. And, see also Pawlowski, 2018, who discusses the planes of satellites not being reproduced in DM simulations.
\par
It is instructive to point out the most prominent ad-hoc `fix' that has been incorporated in these simulations, which is this: Had present-day galaxies retained the cosmic proportion of baryons to DM they started with, condensing from the cosmic `soup',\footnote{Such was the prevalent thought up to the early 1990s, when the calamity of such a scenario was brought home. This was due to the realization that $f_b$ is not about 1:20, as was thought before the entrance of the `dark energy'.} which is $f_b\approx 1/5$, the predicted mass-asymptotic-speed relation (MASR aka BTFR) in \LCDM would be significantly off the observed one, by a factor between 10 and 50. This is shown in Fig. \ref{btfr}, where the \LCDM expectation assuming full retention of the baryons in galaxies is seen to be much higher than the observed relation. This expectation is based on the relation between the mass, $M_h$, of DM halos and the maximum rotational speed they produce, $V_{max}$, and assuming that the baryon mass is $f_b M_h$. The simple fact is that the baryon fraction observed in galaxies today is much smaller than the cosmic value \LCDM endorses today.
\par
To remove these large discrepancies, later simulations introduced an ad-hoc device called `feedback' whereby over the evolution of galaxies, different processes -- such as activity in the galactic nucleus, supernova explosions, and others, have removed from the galaxy almost all the baryons. All these mechanisms cannot be calculated from first principles and are put into simulations as recipes with much freedom as to where in the galaxy, and when, they occur, and to what extent they act in galaxies of different types. (See Posti, \& al., 2019 for possible challenges to this `feedback' device.)
\par
In the first place this freedom was used to reproduce the observed baryon-to-DM mass ratio that matches observations for different galactic masses, in other words to make the MASR go down by a factor of  10-50 to match the observed relation. For example, according to Santos-Santos \& al., 2016:
{\it `Progress was made} (in the simulations) {\it by implementing increasingly effective recipes for feedback from supernovae... and the inclusion of other forms of feedback from massive stars... The benchmark for assessing this progress has primarily been the ability to match the Tully-Fisher relation ..., with recent simulations succeeding at this, and in particular matching the baryonic Tully-Fisher relation (BTFR), for galaxies over a range of masses.'}
(here, BTFR=MASR.)
\par
To recapitulate, `feedback' -- a major adjustment to the DM paradigm -- has not been introduced because of an {\it a priori} need to include the correct underlying physics; it's sole role is to save the paradigm from a disastrous clash with the measurements.
\par
I stress this, because `feedback' is a prime example of what Lakatos (1971) defined as a `degenerating problemshift', which he saw (with us, I should think) as a sign of sickness in a scientific theory (see also the discussion in Sec. \ref{balkan}).
\par
Beyond this major ad-hoc adjustment, much freedom is left for further adjustments in the simulations.

\subsection{MOND-DM hybrids and other compromises\label{hybrids}}
{\it Usually the opponents of a new paradigm can
legitimately claim that even in the area of crisis it is little superior to its
traditional rival. Of course, it... has
disclosed some new regularities. But the older paradigm can presumably
be articulated to meet these challenges... Both
Tycho Brahe's earth-centered astronomical system and the later versions
of the phlogiston ... were quite successful.} (Kuhn, 1962; in `The structure of scientific revolutions'.)

{\it  Since he (the astronomer) cannot in any way attain to the true causes, he will adopt whatever suppositions enable the motions to be computed correctly from the principles of geometry for the future as well as for the past. For these hypotheses (Copernicus's)
need not be true nor even probable. On the contrary, if they provide a calculus consistent with the observations, that alone is enough.} (Osiander' introduction to `De revolutionibus'.)
\par
MOND has made many successful predictions, and works as well as can be expected, in galactic systems. Until recently, it had not had a fully working answer to the cosmological mass discrepancy. (But see the allusion in Sec. \ref{cosmology} to the unpublished work of Skordis \& Z{\l}o\'{s}nik, describing a working MOND cosmology). \LCDM accounts for cosmology and structure formation by invoking DM and `dark energy', and is a good working hypothesis in this realm (with some hints of anomalies; see, e.g., point 7 in Sec. \ref{worries}).
On the other hand it does much poorer than MOND on galactic scales. At this stage none of the two paradigms is generally accepted, nor generally discarded, exactly because each appears to have advantages and shortcomings.
\par
A well-known outcome of such temporary stalemates is the appearances of hybrid pictures that strive to combine the best of the two worlds.
In the case of MOND vs. DM,
such hybrid pictures usually envisage some omnipresent medium that acts, like DM, by its added gravity in the cosmological context. But in galactic systems its main effect is to mediate some new interactions between baryons, dictated so as to reproduce MOND phenomenology. Such are, for example, the suggestion of `polarized DM' (Blanchet 2007, Blanchet \& Le Tiec 2009, Blanchet \& Heisenberg 2015), `Dark Fluid' (Zhao 2008), and
`Superfluid DM' (Khoury 2016, Berezhiani \& Khoury 2016).
\par
A famous historical example of such a compromise is Tycho Brahe's `geoheliocentric' world system mentioned in the quote above -- conflating the Ptolemaic and Copernican systems -- whereby the planets (other than the Earth) revolve around the sun, but the sun, in turn, with its planetary entourage (and the moon and sphere of stars) revolves around the Earth.
It helped retain the benefits of the Copernican picture without some of its perceived drawbacks, and without contradicting, e.g., the strong biblical evidence of the sun standing still in Gibeon, and the cherished centrality of the Earth.
\par
Such an appearance of hybrid models is not the only type of attempt to smooth out conflicts between two opposing paradigms. Another is to assert in effect: `Yes, the new paradigm seems to work better, but this is only an apparent picture `saving the phenomena', a `mathematical tool' that only summarizes more efficiently the results of the old paradigm.
\par
Historical examples of such proposed compromises abound: The well-known assertions to the effect that `The heliocentric, Copernican system is only a more efficient computational tool for calculating planetary motions on the sky, but not how things really are' (as in the above citation from Osiander's preface to the `De revolutionibus').
\par
Another example is the debate about the reality of atoms and molecules, and the compromises (advocated e.g., by Hertz) that atoms are not real, only `images' or `pictures' (`bilder') useful as descriptive tools. Even Boltzmann, whose life work was based on matter being made of atoms and molecules (and who must have believed in their reality) is said to have clung, for some time, to such a compromise in a wish to appease his redoubtable positivist antagonists, such as Mach.
\par
More recently, it was the fate of quarks, the building blocks of hadrons in the standard model of particle physics to be assigned, by some, the role of mere `computational tools'.
\par
In connection with MOND we often hear statements to the effect that `Yes, MOND works well on galactic scales; but this is only an efficient summary of how DM works, somehow.' (starting with Kaplinghat \& Turner, 2002, rebutted in Milgrom, 2002, and continuing to this day).
\subsection{MOND-DM discriminants\label{discriminants}}
There are consequences of MOND that cannot be mimicked with DM -- at least not with the kind of DM that is assumed at present.
In principle, these discriminating predictions can be used to decide between MOND and DM, but so far their study has not lead to generally accepted, definite conclusions, because some of them are still beyond reach, and for others because the interpretation of the results is not clear cut.
\par
I list briefly some of these, without much comment, and refer the reader to works on these issues.
\par
Negative phantom matter: From the definition of the phantom density in Sec. \ref{mimic},
it is not guaranteed that $\r_p$ is non-negative. In fact, it has been shown (e.g.,
Milgrom, 1986) that in some MOND theories one can identify locations where $\r_p<0$, and a DM interpretation will have to contend with negative masses.
\par
Dynamical friction: The DM medium around galactic systems should exert `dynamical friction' on objects moving through it. Such effects do not occur in MOND in the same way, since only the baryons are there to exert friction. Some studies that attempt to identify such effects can be found e.g., in Nipoti \& al., 2008; Angus \& al., 2011; and
Oehm \& al., 2017.
\par
Binary stars in the Milky Way: If we can measure the orbits in very wide binaries in the Milky Way, so wide that their intrinsic accelerations are at or below $\az$, we might find MOND departures from Newtonian dynamics, not expected in the DM paradigm (see, e.g.,
Hernandez \& al., 2019; Pittordis \& Sutherland, 2019; Banik \& Zhao, 2018; Banik \& Kroupa, 2019).
\par
The external-field effect (Milgrom 1983): In MOND, the external gravitational acceleration field in which a system is falling can greatly affect the system's internal dynamics. This MOND effect has no analog in Newtonian dynamics (with or without DM). If clearly identified, it will strongly point to MOND.
\par
Some possible identifications of this effect at work have been discussed, e.g., in McGaugh \& Milgrom (2013a, 2013b) (for dwarf satellites of Andromeda); McGaugh, 2016 (for the ultra-faint-and-large dwarf, Crater II); Haghi \& al., 2019; and Kroupa \& al., 2019.
\par
Disc stability: The halo of DM that accounts for the rotation-curve discrepancy in disc galaxies would also have stability effects on the disc itself, which are different from those in MOND (Milgrom, 1989a; Brada \& Milgrom, 1999a; Tiret \& Combes, 2008; Banik \& Zhao, 2018;  Banik \& al., 2018).
\par
Features on rotation curves: In some galaxies there are features in the observed mass distribution of the baryons; these can be undulations, sharp rises, or drops, etc. These lead to corresponding features in the gravitational field produced by the baryons, and hence in the Newtonian rotation curve. In MOND, baryons are the only source of gravity; so the MOND rotation curve clearly retains such features.
\par
In contradistinction, the \LCDM halos are not expected to have pronounced features in their radial density distribution. Even if they did, these would be only weakly reflected in the halo gravitational field, and they would, anyway, not be expected to occur where the baryon features are.
\par
In high-acceleration regimes, where there supposedly is little contribution to gravity from the DM halo, such correlated features in the baryon gravitational field and the total observed rotation curve are understood as a pure-baryon effect. However, we also see examples of such correspondences in regions where the mass discrepancy is large. There, the gravity of the purported DM halo is dominant, and thus we expect such baryon-produced features to be smoothed out.
\par
Imagine, for example, that the Newtonian, baryonic rotation speed at some radius is $V_b$, and exhibits a small undulation of amplitude $\d V_b$ in a small range of radii, and that this occurs where the measured rotation $V\gg V_b$ (i.e., where the mass discrepancy is large).
Then, the relative undulation in the observed velocity in the DM paradigm is expected to be,
$(\d V/V)\approx (V_b/V)^2(\d V_b/V_b)$,\footnote{Because $V=(V_b^2+V_h^2)^{1/2}$, where $V_h\gg V_b$ is the velocity produced by the putative halo alone ($\d V_h=0$ for a smooth halo). Thus $\d V\approx V_b\d V_b/V$.} whereas in MOND we expect $(\d V/V)\approx (1/2)(\d V_b/V_b)$. Namely, the reduction in MOND is only by a factor $1/2$, no matter how large the mass discrepancy is, compared with the possibly much more severe reduction by a factor of $(V_b/V)^2$ in DM.
\par
\begin{figure}[!ht]
\includegraphics[width=1.\columnwidth]{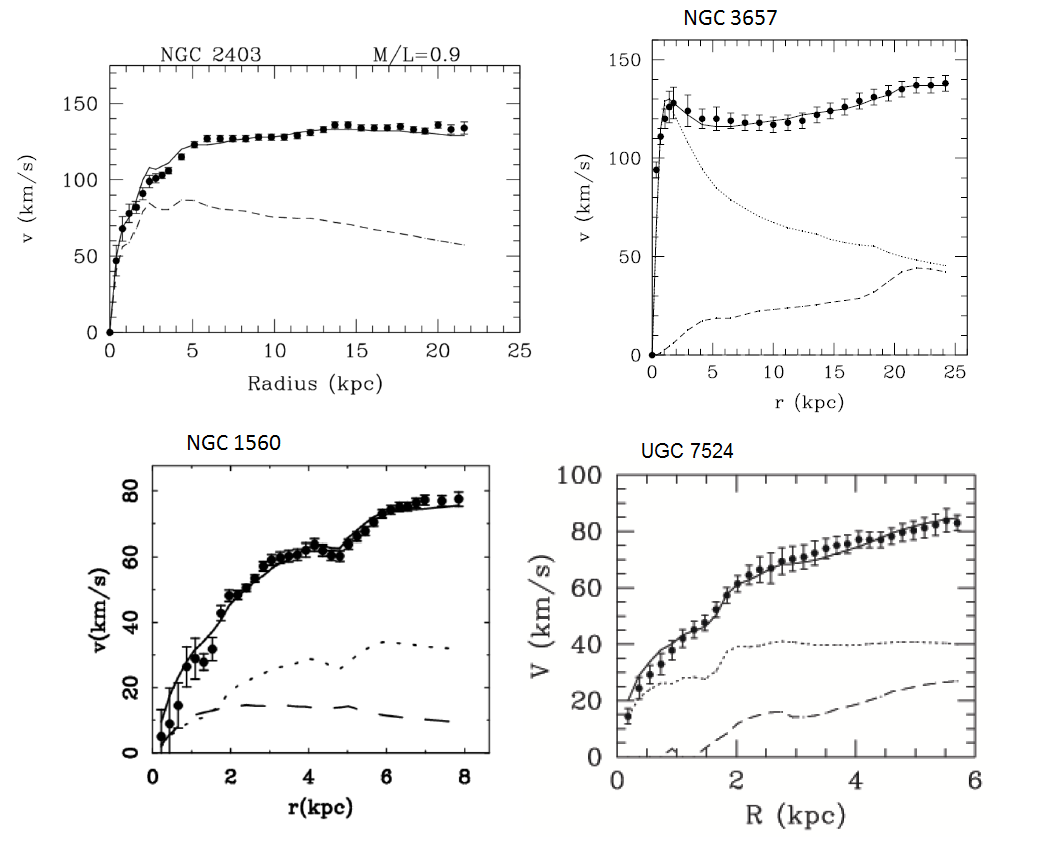}
\caption{Examples of rotation curves with features that correspond to features in the disc mass distribution (as reflected in the Newtonian rotation curves calculated without DM, and shown here for the gas and stellar components as the lower curves in each frame). These features are reproduced in the rotation curves predicted by MOND (the curves going through the data points).
From Sanders \& McGaugh, 2002, Sanders, 2007, and Sanders (private communication). \label{rcs}}
\end{figure}

Figure \ref{rcs} shows a few examples, and several more can be found in Figs. \ref{sparc}  and \ref{sanders19}. See also discussions of this important point e.g., in Begeman \& al., 1991; Gentile \& al., 2010; McGaugh, 2014; and Sanders, 2016.\footnote{Rotation-curve features can also result as artefacts, and do not always correspond to features in the gravitational field; so care is needed in their interpretation.}

\section{Where on the MOND timeline are we?\label{timeline}}
{\it ``To be sure, the introduction of the quantum of action has not yet produced a genuine quantum theory. In fact, the path the research worker must yet tread to it is not less than that from the discovery of the velocity of light by Olaf R\"{o}mer to the establishment of Maxwell’s theory of light.''} (Planck, 1920; Nobel-Prize lecture).

Planck's Nobel lecture, twenty years after the advent of the quantum, was a good time to take stock of that paradigm. Much has been achieved by then -- enough to indicate strongly that the quantum was the right track. But there was still much to be done, as Planck explained using a historical analogy himself, and as we know now with the benefit of hindsight. (The ending paragraph of this paper, also a quote from Planck, 1920, is a fitting complement to the above quote.)
\par
Arguably, MOND is now in a more advanced state than is described for quantum mechanics in the above quote from Planck. We are beyond the stage where we have a collection of phenomena and phenomenological laws where the same constant $\az$ appears. As explained in Sec. \ref{theory}, and shown in Fig. \ref{flow}, we do have full theories that yield those laws as corollaries, and that can be used to calculate the dynamics of whole systems. But as also explained in Sec. \ref{theory}, we are not fully satisfied with the existing MOND theories, and there are still observations (especially in cosmology) that MOND does not account for.
\par
If we are looking for a period in the development of quantum mechanics that is roughly analogous to where MOND is today, it seems to me that this would be the period right after the advent of the Schr\"{o}dinger equation and its first uses; e.g., in reproducing the spectrum of the hydrogen atom. We already had in place the role of $\hbar$, the discreteness of states, the de Broglie concept of the wave nature of particles, and the relation between momentum and wavelength. We had already Heisenberg's matrix mechanics and its equivalence to Schr\"{o}dinger's theory. The role of spin and the Pauli exclusion principle were just being clarified. But there was sill much left to be done. There were still to come the correct treatment of multi-electron atoms, the understanding of the meaning of the Schr\"{o}dinger wave function (debated still today), the understanding of measurements, the Dirac equation, relativistic, quantum field theory, and the still-elusive quantum gravity.
\par
In comparison with GR, it can be argued that the amount of data confronted successfully with MOND by now far exceeds those for GR even today. But from the point of view of theoretical development and understanding of fundamentals, MOND is not yet on par with GR.
From the theoretical point of view we are perhaps in a state similar to that several years prior to the advent of GR (say 1910-1914):
By this time we already had special relativity; so the need for an extension of Newtonian gravity that is compatible with special relativity was pressing. Several prominent physicists were publishing on the subject, such as Nordstr\"{o}m, Abraham, Mie, and Einstein himself with collaborators, such as Fokker and Grossmann (see, e.g., Einstein \& Grossmann, 1914).
\par
Einstein's equivalence principle was put forth as a basic guiding tenet for the desired theory, and it was already used to derive some phenomenological laws: light bending and gravitational redshift (the predicted magnitude of the former was famously wrong by a factor of two). There were, in fact, already proposed several theories that satisfied the basic tenets of being compatible with special relativity, and giving Newtonian gravity in the nonrelativistic limit.
There was, for example, Nordstr\"{o}m's `first theory', then Nordstr\"{o}m's `second theory' (see Norton, 2005, for the history of these theories, and Deruelle, 2011 for more on their physics): It was shown in
Einstein \& Fokker, 1914 that this latter theory can be cast as a  metric theory with the important gravity-is-geometry element of GR already in place. It is compatible with special relativity and has Newtonian gravity as its nonrelativistic limit, and is derived from an action. Furthermore, it satisfies the strong equivalence principle (e.g., Deruelle, 2011).
\par
However, Nordstr\"{o}m's theory had to be abandoned, and give way to GR, another theory satisfying the basic tenets, on phenomenological grounds: It predicted the wrong anomalous perihelion shift of the plant Mercury (already known at the time), and it predicts no light bending, which turned out to be in conflict with the later observation of such bending.
\par
This pre-GR period is also characterized by tortuous progress as regards relativistic-gravity theories [e.g., Pais, 1982 (chapters 11-13), and Weinstein 2012, 2019], similar to what we are experiencing today with MOND theory.

\par
I end with the optimistic and foresighted quote from Planck, 1920, in the face of the difficulties still besetting the quantum paradigm at the time:
{\it ``Be that as it may, in any case no doubt can arise that science will master the dilemma, serious as it is, and that which appears today so unsatisfactory will in fact eventually, seen from a higher vantage point, be distinguished by its special harmony and simplicity. Until this aim is achieved, the problem of the quantum of action will not cease to inspire research and fructify it, and the greater the difficulties which oppose its solution, the more significant it finally will show itself to be for the broadening and deepening of our whole knowledge in physics.''}


\begin{thebibliography}{}
\bibitem{vanalbada85}van Albada,  \& al. (1985). Distribution of dark matter in the spiral galaxy NGC 3198.
    Astrophys. J. 295, 305-313.
\bibitem{angus08}Angus, G.W. (2008). Dwarf spheroidals in MOND.
    Mon. Not. Roy. Astron. Soc. 387, 1481-1488.
\bibitem{angus11}Angus, G.W., Diaferio, A., \& Kroupa, P. (2011). Using dwarf satellite proper motions to determine their origin. Mon. Not. Roy. Astron. Soc. 416, 1401-1409.
\bibitem{bz18}Banik, I., \& Zhao, H.S. (2018). Testing gravity with wide binary stars like $\alpha$ Centauri.
    Mon. Not. Roy. Astron. Soc. 480, 2660-2688.
\bibitem{bk19}Banik, I., \& Kroupa, P. (2019).Directly testing gravity with Proxima Centauri.
    Mon. Not. Roy. Astron. Soc. 487, 1653-1661.
\bibitem{bmz18}Banik, I., Milgrom, M., \& Zhao, H.S. (2018).
Toomre stability of disk galaxies in quasi-linear MOND.
   arXiv:1808.10545
\bibitem{barnes07}Barnes, E.I., Kosowsky, A., \& Sellwood, J.A. (2007).
	Milgrom Relation Models for Spiral Galaxies from Two-dimensional Velocity Maps. Astron. J. 133, 1698-1709.
\bibitem{begeman91}Begeman, K.G.,  Broeils, A.H., \& Sanders, R.H. (1991). Extended rotation curves of spiral galaxies - Dark haloes and modified dynamics.  Mon. Not. R. Astron. Soc.   249,  523.
\bibitem{bekenstein04}J.D. Bekenstein, J.D. (2004). Relativistic gravitation theory for the modified Newtonian dynamics paradigm. Phys. Rev. D 70, 083509.
\bibitem{bk16}Berezhiani, L. \& Khoury, J. (2016). Dark matter superfluidity and galactic dynamics. Phys. Lett. B   753,  639.
\bibitem{bh18}Bertone, G., \& Hooper, D. (2018). History of dark matter. Rev. Mod. Phys. 90, 045002
\bibitem{bh18}Bertone, G., \& Tait, T.M.P. (2018). A new era in the search for dark matter. Nature 562, 51-56.
\bibitem{bilek2019a}B\'{i}lek, M., Samurovi\'{c}, S., \& Renaud, F.  (2019a). Study of gravitational fields and globular cluster systems of early-type galaxies.
    Astron. Astrophys. 625, A32.
\bibitem{bilek2019b}B\'{i}lek, M., Samurovi\'{c}, S., \& Renaud, F.  (2019b). Imprint of the galactic acceleration scale on globular cluster systems. Astron. Astrophys. L., in press, arXiv:1908.04783
\bibitem{blanchet07}Blanchet, L. (2007). Dipolar particles in general relativity.  Class. Quant. Grav.   24,  3541.
\bibitem{blh17}Blanchet, L. \&  and Heisenberg, L. (2017). Dark matter via massive bigravity. Phys. Rev. D  96  083512.
\bibitem{blt09}Blanchet, L \& Le Tiec, A.  (2009). Dipolar dark matter and dark energy. Phys. Rev. D 80,  023524.
\bibitem{bm11}Blanchet, L. \& Marsat, S. (2011). Modified gravity approach based on a preferred time foliation. Phys. Rev. D 84, 044056.
\bibitem{bosma78}Bosma, A. (1978). The distribution and kinematics of neutral hydrogen in spiral galaxies of various morphological types. PhD thesis, University of Groningen.
\bibitem{bosma81}Bosma, A. (1981). 21-cm line studies of spiral galaxies. II. The distribution and kinematics of neutral hydrogen in spiral galaxies of various morphological types. Astron. J. 86, 1825-1846.
\bibitem{brada99}Brada, R. \& Milgrom, M. (1999). The Modified Newtonian Dynamics Predicts an Absolute Maximum to the Acceleration Produced by ``Dark Halos''. Astrophys J. lett. 512, L17-L18.
\bibitem{brada99a}Brada, R. \& Milgrom, M. (1999a). Stability of Disk Galaxies in the Modified Dynamics.  Astrophys J. 519, 590-598.
\bibitem{brimioulle13}Brimioulle, F. \& al. (2013). Dark matter halo properties from galaxy-galaxy lensing. Mon. Not. R. Astron. Soc. 432, 1046.
\bibitem{brouwer17}Brouwer, M.M.  \& al. (2017). First test of Verlinde's theory of emergent gravity using weak gravitational lensing measurements. Mon. Not. R. Astron. Soc. 466, 2547.
\bibitem{bb17}Bullock, J.S. \& Boylan-Kolchin, M. (2017). Small-Scale Challenges to the \LCDM Paradigm.
    Ann. Rev. Astron. Astrophys. 55, 343-387.
\bibitem{buote19}Buote, D.A. \& Barth, A.J. (2019). The Extremely High Dark Matter Halo Concentration of the Relic Compact Elliptical Galaxy Mrk 1216. Astrophys. J. 877, 91.
\bibitem{cz10}Clifton, T. \& Zlosnik, T.G. (2010). FRW cosmology in Milgrom's bimetric theory of gravity. Phys. Rev. D  81,  103525.
\bibitem{clifton12}Clifton, T. \& al. (2012). Modified gravity and cosmology. Phys. Rep. 513, 1-189.
\bibitem{clowe06}Clowe, D. \& al. (2006). A Direct Empirical Proof of the Existence of Dark Matter. Astrophys. J. lett. 648, L109-L113.
\bibitem{dbm98}de Blok, W.J.G., \& McGaugh, S.S. (1998). Testing Modified Newtonian Dynamics with Low Surface Brightness Galaxies: Rotation Curve FITS. Astrophys. J. 508, 132-140.
\bibitem{deffayet14}Deffayet, C., Esposito-Farese, G., \&  Woodard, R.P. (2014).  Field equations and cosmology for a class of nonlocal metric models of MOND. Phys. Rev. D 90, 064038.
\bibitem{popolo17}Del Popolo, A. \& Le Delliou, M. (2017). Small Scale Problems of the \LCDM Model: A Short Review.  Galaxies 5, 17.
\bibitem{deruelle11}Deruelle, N. (2011). Nordstr\"{o}m's scalar theory of gravity and the equivalence principle.
    Gen. Rel. Grav. 43, 3337-3354.
\bibitem{donato09}Donato, F. \& al. (2009). A constant dark matter halo surface density in galaxies. Mon. Not. Roy. Astron. Soc. 397, 1169-1176.
\bibitem{efstathiou90}Efstathiou, G., Sutherland, W.J., \& Maddox, S.J. (1990). The cosmological constant and cold dark matter.
Nature 348, 705-707.
\bibitem{ef14}Einstein, A. \& Fokker, A.D. (1914). Die Nordstr\"{o}msche Gravitationstheorie vom Standpunkt des absoluten Differentialkalk\"{u}ls. Ann. d. Phys. 349, 321-328.
\bibitem{eg14}Einstein, A. \& Grossmann, M. (1914). Kovarianzeigenschaften der Feldgleichungen der auf die verallgemeinerte Relativit\"{a}tstheorie gegr\"{u}ndeten Gravitationstheorie. Zeitschr. f. Mathem. u.  Phys. 63, 215-225.
\bibitem{ettori19}Ettori, S., \& al. (2019). Hydrostatic mass profiles in X-COP galaxy clusters.  Astron. Astrophys. 621, A39.
\bibitem{fm12}Famaey, B. \& McGaugh, S. (2012).  Modified Newtonian Dynamics (MOND): Observational Phenomenology and Relativistic Extensions. Liv. Rev. Rel.  15,  10.
\bibitem{feng10}Feng, J.L. (2010). Dark Matter Candidates from Particle Physics and Methods of Detection. Ann. Rev. Astron. Astrophys. 48, 495-545.
\bibitem{franck16}Franck, J.R. \& McGaugh, S.S. (2016). The Candidate Cluster and Protocluster Catalog (CCPC) II. Spectroscopically Identified Structures Spanning $2<z<6.6$. Astrophys. J. 833, 15.
\bibitem{gentile10}Gentile, G., Baes, M., Famaey, B., \& van Acoleyen, K. (2010). Mass models from high-resolution HI data of the dwarf galaxy NGC 1560.
     Mon. Not. Roy. Astron. Soc. 406, 2493-2503.
\bibitem{gentile11}Gentile, G., Famaey, B., de Blok, W.J.G. (2011). THINGS about MOND. Astron. \& Astrophys. 527, A76-86.
\bibitem{haghi19}Haghi, H., \& al. (2019). A new formulation of the external field effect in MOND and numerical simulations of ultra-diffuse dwarf galaxies - application to NGC 1052-DF2 and NGC 1052-DF4.
    Mon. Not. Roy. Astron. Soc. 487, 2441-2454.
\bibitem{hashimoto18}Hashimoto, T., \& al. (2018). The onset of star formation 250 million years after the Big Bang. Nature 557, 392–395.
\bibitem{hees16}Hees, A., \& al. (2016). Combined Solar system and rotation curve constraints on MOND. Mon. Not. Roy. Astron. Soc. 455, 449-461.
\bibitem{hernandez19}Hernandez, X., Cortés, R.A.M., Allen, C., and Scarpa, R. (2019). Challenging a Newtonian prediction through Gaia wide binaries.
    Inter. J. Mod. Phys. D  28, 1950101.
\bibitem{ho10}Ho, C.M., Minic, D., \& Ng, Y.J. (2010). Cold dark matter with MOND scaling. Phys. Lett. B   693  567 (2010).
\bibitem{humphrey11}Humphrey, P.J. \& al. (2011). A Census of Baryons and Dark Matter in an Isolated, Milky Way Sized Elliptical Galaxy. Astrophys. J. 729, 53.
\bibitem{humphrey12}Humphrey, P.J. \& al. (2012). The ElIXr Galaxy Survey. II. Baryons and Dark Matter in an Isolated Elliptical Galaxy. Astrophys. J. 755, 166.
\bibitem{kaplinghat02}Kaplinghat, M. \& Turner, M. (2002). How Cold Dark Matter Theory Explains Milgrom's Law. Astrophys. J. Lett. 569, L19-L22.
\bibitem{keller17}Keller, B.W. \& Wadsley, J. (2017). \LCDM is Consistent with SPARC Radial Acceleration Relation. Astrophys. J. Lett. 835, L17-L21.
\bibitem{khoury16}Khoury, J. (2016). Another path for the emergence of modified galactic dynamics from dark matter superfluidity. Phys. Rev. D   93,  103533.
\bibitem{kt11}Kiselev, V.V. \& Timofeev, S.A. (2011). The Holographic Screen at Low Temperatures. Mod. Phys. Lett. A26, 109.
\bibitem{kk11}Klinkhamer, F.R. \& Kopp, M. (2011). Entropic Gravity, Minimum Temperature, and Modified Newtonian Dynamics. Mod. Phys. Lett. A   26  2783.
\bibitem{kroupa15}Kroupa, P. (2015). Galaxies as simple dynamical systems: observational data disfavor dark matter and stochastic star formation. Canad. J. Phys. 93, 169-202.
\bibitem{kroupa16}Kroupa, P. (2016). The observed spatial distribution of matter on scales ranging from 100kpc to 1Gpc is inconsistent with the standard dark-matter-based cosmological models.
in "Cosmology on Small Scales 2016", M. Krizek and Y. Dumin (Eds.), Institute of Mathematics, Czech Academy of Sciences, Prague: ISBN 978-80-85823-66-0
arXiv:1610.03854.
\bibitem{kroupa19}Kroupa, P., \& al. (2019). Does the galaxy NGC1052-DF2 falsify Milgromian dynamics? Nature (BCA) 561, E4–E5.
\bibitem{kuhn62}Kuhn, T. (1962). The Structure of Scientific Revolutions. University of ChicagoPress
\bibitem{lahavmassimi14}Lahav, O. \& Massimi, M. (2014). Dark energy, paradigm shifts, and the role of evidence.
    Astron. Geophys. 55, 3.12-3.15
\bibitem{lakatos71}Lakatos, I. (1971). History of science and its rational reconstructions (1971). In: R. Buck, \& R. S. Cohen (Eds.), Boston Studies in the Philosophy of Science, Vol. VIII). Dordrecht: Reidel.
\bibitem{lazutkina17}Lazutkina, A. (2017). Theoretical terms of contemporary cosmology as intellectual artifacts.  In J\"{u}rgen H. Franz \& Karsten Berr (eds.) Welt der  Artefakte. Berlin: Frank \& Timme. 63 – 70, arXiv:1707.05235.
\bibitem{lelli16}Lelli, F. \& al. (2016). The Relation between Stellar and Dynamical Surface Densities in the Central Regions of Disk Galaxies. Astrophys. J. Lett. 827, 19L (2016).
\bibitem{li18}Li, P., Lelli, F., McGaugh, S., \& Schombert, J. (2018). Fitting the radial acceleration relation to individual SPARC galaxies. Astron. Astrophys. 615, A3.
\bibitem{lc11}Li, X. \& Chang, Z. (2011). Debye Entropic Force and Modified Newtonian Dynamics. Commun. Theor. Phys.   55  733.
\bibitem{makkar11}Makarov, D. \& Karachentsev, I. (2011).  Galaxy groups and clouds in the local ($\sim 0.01$) Universe.  Mon. Not. R. Astron. Soc. 412, 2498.
\bibitem{massimi18}Massimi, M. (2018). Three problems about multi-scale modelling in cosmology. Stud. Hist. Philos. Mod. Phys. 64, 26-38.
\bibitem{maxwell92}Maxwell, J.C. (1892). A Treatise on Electricity and Magnetism, (article 771). NY, Dover Publications 1954 (reprinted from the 1891 edition, Clarendon Press, Oxford, UK)
\bibitem{mcgaugh99}McGaugh, S.S. (1999). How Galaxies Don't Form: The Effective Force Law in Disk Galaxies.
 ASP Conference Series 182 (Eds. Merritt, Valluri, \& Sellwood).
\bibitem{mcgaugh11}McGaugh, S.S. (2011). Novel Test of Modified Newtonian Dynamics with Gas Rich Galaxies. Phys. Rev. Lett. 106, 121303.
\bibitem{mcgaugh12}McGaugh, S.S. (2012). The Baryonic Tully-Fisher Relation of Gas Rich Galaxies as a Test of \LCDM and MOND.
 Astron. J. 143, 40-54.
\bibitem{mcgaugh14}McGaugh, S.S. (2014). The Third Law of Galactic Rotation. Galaxies,  2, 601-622.
\bibitem{mcgaugh15}McGaugh, S.S. (2015). A Tale of Two Paradigms: the Mutual Incommensurability of \LCDM and MOND. Canad. J. Phys. 93, 250-259.
\bibitem{mcgaugh16}McGaugh, S.S. (2016).   MOND Prediction for the Velocity Dispersion of the “Feeble Giant” Crater II. Astrophys. J. Lett. 832, L8.
\bibitem{mcgaugh18}McGaugh, S.S. (2018). Predictions for the Sky-Averaged Depth of the 21 cm Absorption Signal at High Redshift in Cosmologies with and without Nonbaryonic Cold Dark Matter . Phys. Rev. Lett. 121, 081305.
\bibitem{mls16}McGaugh, S.S., Lelli, F., \& Schombert, J.M. (2016). Radial Acceleration Relation in Rotationally Supported Galaxies.  Phys. Rev. Lett. 117, 201101.
\bibitem{mm13a}McGaugh, S. \& Milgrom, M. (2013a). Andromeda Dwarfs in Light of Modified Newtonian Dynamics. Astrophys. J. 766, 22.
\bibitem{mm13b}McGaugh, S. \& Milgrom, M. (2013b). Andromeda Dwarfs in Light of MOND. II. Testing Prior Predictions. Astrophys. J. 775, 139.
\bibitem{mw10}McGaugh, S.S. \& Wolf, J. (2010). Local Group Dwarf Spheroidals: Correlated Deviations from the Baryonic Tully-Fisher Relation. Astrophys. J. 722, 248-261.
\bibitem{merritt16}Merritt, D. (2016). Cosmology and convention. Stud. Hist. Philos. Mod. Phys. 57, 41-52.
\bibitem{merritt19}Merritt, D. (2020). {\it A Philosophical Approach to MOND: Assessing the Milgromian Research Program in Cosmology}, Cambridge University Press.
\bibitem{milgrom83}Milgrom, M. (1983). A modification of the Newtonian dynamics as a possible alternative to the hidden mass hypothesis.  Astrophys. J.   270,  365.
\bibitem{milgrom83a}Milgrom, M. (1983a). A modification of the Newtonian dynamics - Implications for galaxies.  Astrophys. J.   270,  371.
\bibitem{milgrom84}Milgrom, M. (1984). Isothermal spheres in the modified dynamics. Astrophys. J. 287, 571-576.
\bibitem{milgrom86}Milgrom, M. (1986). Can the Hidden Mass Be Negative? Astrophys. J. 306, 9.
\bibitem{milgrom88}Milgrom, M. (1988). On the Use of Galaxy Rotation Curves to Test the Modified Dynamics. Astrophys. J. 333, 689.
\bibitem{milgrom89}Milgrom, M. (1989).	Alternatives to Dark Matter. Comm. Astrophys. 13 (4), 215.
\bibitem{milgrom89a}Milgrom, M. (1989a). On Stability of Galactic Disks in the Modified Dynamics and the Distribution of Their Mean Surface-Brightness.
    Astrophys. J. 338, 121.
\bibitem{milgrom94}Milgrom, M. (1994). Dynamics with a Nonstandard Inertia-Acceleration Relation: An Alternative to Dark Matter in Galactic Systems. Ann. Phys. 229, 384.
\bibitem{milgrom99}Milgrom, M. (1999). The modified dynamics as a vacuum effect. Phys. Lett. A   253,  273.
\bibitem{milgrom02}Milgrom, M. (2002). Do Modified Newtonian Dynamics Follow from the Cold Dark Matter Paradigm?     Astrophys. J. Lett. 571, L81-L83.
\bibitem{milgrom08}Milgrom, M. (2008). Marriage à-la-MOND: Baryonic dark matter in galaxy clusters and the cooling flow puzzle. New Astron. Rev. 51, 906-915.
\bibitem{milgrom09}Milgrom, M. (2009). The Mond Limit from Spacetime Scale Invariance. Astrophys. J.   698,  1630.
\bibitem{milgrom09a}Milgrom, M. (2009a). The central surface density of `dark haloes' predicted by MOND. Mon. Not. R. Astron. Soc.   398  1023-1026.
\bibitem{milgrom09b}Milgrom, M. (2009b). Bimetric MOND gravity. Phys. Rev. D 80, 123536.
\bibitem{milgrom10}Milgrom, M. (2010). Cosmological fluctuation growth in bimetric MOND. Phys. Rev. D 82, 043523.
\bibitem{milgrom12}Milgrom, M. (2012). Testing MOND over a Wide Acceleration Range in X-Ray Ellipticals. Phys. Rev. Lett. 109, 131101.
\bibitem{milgrom13}Milgrom, M. (2013). Testing the MOND Paradigm of Modified Dynamics with Galaxy-Galaxy Gravitational Lensing. Phys. Rev. Lett. 111, 041105.
\bibitem{milgrom14}Milgrom, M. (2014). The MOND paradigm of modified dynamics.  Scholarpedia  9(6)  31410.
\bibitem{milgrom14a}Milgrom, M. (2014a). MOND laws of galactic dynamics. Mon. Not. R. Astron. Soc.   437  2531.
\bibitem{milgrom14b}Milgrom, M. (2014b). General virial theorem for modified-gravity MOND.
Phys. Rev. D 89, 024016.
\bibitem{milgrom15}Milgrom, M. (2015). MOND theory.  Canad. J. Phys., 93, 107-118.
\bibitem{milgrom15a}Milgrom, M. (2015a). Road to MOND: A novel perspective. Phys. Rev. D 92, 044014.
\bibitem{milgrom16}Milgrom, M. (2016). Universal Modified Newtonian Dynamics Relation between the Baryonic and "Dynamical" Central Surface Densities of Disc Galaxies. Phys. Rev. Lett. 117, 141101.
\bibitem{milgrom16a}Milgrom, M. (2016a). MOND impact on and of the recently updated mass-discrepancy-acceleration relation. arXiv:1609.06642.
\bibitem{milgrom16b}Milgrom, M. (2016b). The \LCDM simulations of Keller and Wadsley do not account for the MOND mass-discrepancy-acceleration relation. arXiv:1610.07538.
\bibitem{milgrom19}Milgrom, M. (2019). MOND in galaxy groups: A superior sample. Phys. Rev. D 99, 044041.
\bibitem{milgrom19a}Milgrom, M. (2019a). MOND from a brane-world picture. In `Jacob Bekenstein Memorial Volume'
 L. Brink, V. Mukhanov, E. Rabinovici, \& and K.K. Phua (Eds.), arXiv:1804.05840.
\bibitem{milgrom19b}Milgrom, M. (2019b). Noncovariance at low accelerations as a route to MOND. Phys. Rev. D 100, 084039.
\bibitem{milgrom16b}Milgrom, M. (2020). The $\az$ -- cosmology connection in MOND. arXiv:2001.09729.
\bibitem{mb87}Milgrom, M. \& Bekenstein, J. (1987). The modified Newtonian dynamics as an alternative to hidden matter.
	IN: Dark matter in the universe; Proceedings of the IAU Symposium, 1985. Dordrecht, D. Reidel  319-330.
\bibitem{ms05}Milgrom, M. \& Sanders, R.H. (2005). MOND predictions of `halo' phenomenology in disc galaxies.
    Mon. Not. Roy. Astron. Soc. 357, 45-48.
\bibitem{nipoti08}Nipoti, C., Ciotti, L., Binney, J., \& Londrillo, P. (2008). Dynamical friction in modified Newtonian dynamics. Mon. Not. Roy. Astron. Soc. 386, 2194-2198.
\bibitem{nv07}Noordermeer, E., \& Verheijen, M. A. W. (2007). The high-mass end of the Tully-Fisher relation. Mon. Not. Roy. Astron. Soc. 381, 1463-1472.
\bibitem{norton05}Norton, J.D. (2005). Einstein, Nordstr\"{o}m, and the early demise of scalar, Lorentz
covariant theories of gravitation. In `The Genesis of General Relativity Vol. 3: Theories of Gravitation in the Twilight of Classical Physics.Part I.', Jrgen Renn (ed.), Kluwer Academic Publisher. http://pitt.edu/*jdnorton/papers/Nordstroem.pdf
(replace in the address * by $\sim$).
\bibitem{oehm17}Oehm, W., Thies, I., \& Kroupa, P. (2017).
Constraints on the dynamical evolution of the galaxy group M81.
    Mon. Not. Roy. Astron. Soc. 467, 273-289.
\bibitem{pais82}Pais, A. (1982). Subtle is the Lord... . Oxford University Press.
\bibitem{papastergis16}Papastergis, E., Adams, E.A.K., \& van der Hulst, J.M. (2016). An accurate measurement of the baryonic Tully-Fisher relation with heavily gas-dominated ALFALFA galaxies. Astron. Astrophys. 593, A39.
\bibitem{pawlowski18}Pawlowski, M.S. (2018). The planes of satellite galaxies problem, suggested solutions, and open questions. Mod. Phys. Lett. A, . 33, 1830004.
\bibitem{pm14}Pawlowski, M.S. \&  McGaugh, S.S. (2014). Perseus I and the NGC 3109 association in the context of the Local Group dwarf galaxy structures. Mon. Not. R. Astr. Soc. 440, 908.
\bibitem{pa12}Pazy, E. \& Argaman, N. (2012). Quantum particle statistics on the holographic screen leads to modified Newtonian dynamics. Phys. Rev. D   85  104021.
\bibitem{peebles17}Peebles, P.J.E. (2017). Growth of the nonbaryonic dark matter theory. Nature Astronomy,  1, 57.
\bibitem{peebles10}Peebles, P.J.E. \& Nusser, A. (2010). Nearby galaxies as pointers to a better theory of cosmic evolution. Nature 465, 565-569.
\bibitem{pikhitsa10}Pikhitsa, P.V. (2010). MOND reveals the thermodynamics of gravity. arXiv1010.0318 (2010).
\bibitem{pittordis19}Pittordis, C. \& Sutherland, W. (2019). Testing Modified Gravity with Wide Binaries in GAIA DR2. arXiv:1905.09619.
\bibitem{planck20}Planck, M. (1920). The Genesis and Present State of Development of the Quantum Theory. Nobel Lecture. The Nobel Prize website: https://www.nobelprize.org/prizes/physics/1918

    /planck/lecture/

\bibitem{posti19}Posti, L., \& al. (2019). Galaxy disc scaling relations: A tight linear galaxy -- halo connection challenges abundance matching. arXiv:1909.01344 (Astron. \& Atrophys., in press).
\bibitem{read19}Read, J.I., Walker, M.G., Steger, P. (2019). Dark matter heats up in dwarf galaxies. Mon. Not. Roy. Astron. Soc. 484, 1401-1420.
\bibitem{rubin80}Rubin, V.C., Ford, W.K., Jr., \& Thonnard, N. (1980). Rotational properties of 21 SC galaxies with a large range of luminosities and radii, from NGC 4605 (R=4kpc) to UGC 2885 (R=122kpc).
    Astrophys. J. 238, 471-487.
\bibitem{sanders90}Sanders, R.H. (1990). Mass discrepancies in galaxies - Dark matter and alternatives. Astron. Astrophys. Rev.  2, 1-28.
\bibitem{sanders96}Sanders, R.H. (1996). The Published Extended Rotation Curves of Spiral Galaxies: Confrontation with Modified Dynamics. Astrophy. J. 473,117.
\bibitem{sanders97}Sanders, R.H. (1997). A Stratified Framework for Scalar-Tensor Theories of Modified Dynamics. Astrophys. J. 480, 492-502.
\bibitem{sanders99}Sanders, R.H. (1999). The Virial Discrepancy in Clusters of Galaxies in the Context of Modified Newtonian Dynamics. Astrophy. J. Lett. 512, L23-L26.
\bibitem{sanders07}Sanders, R.H. (2007). Modified Gravity Without Dark Matter. In `The Invisible Universe: Dark Matter and Dark Energy, Lecture Notes in Physics, Volume 720. Springer-Verlag Berlin Heidelberg, 375.
\bibitem{sanders10}Sanders, R.H. (2010). The Dark Matter Problem: A Historical perspective. Cambridge University Press.
\bibitem{sanders11}Sanders, R.H. (2011). Hiding Lorentz invariance violation with MOND. Phys. Rev. D 84, 084024 (2011).
\bibitem{sanders16}Sanders, R.H. (2016). Deconstructing Cosmology. Cambridge University Press.
\bibitem{sanders19}Sanders, R.H. (2019). The prediction of rotation curves in gas-dominated dwarf galaxies with modified dynamics. Mon. Not. Roy. Astron. Soc. 485, 513-521.
\bibitem{sanmc02}Sanders, R.H. \& McGaugh, S.S. (2002). Modified Newtonian Dynamics as an Alternative to Dark Matter. Ann. Rev.  Astron. Astrophysi. 40, 263-317.
\bibitem{santos16}Santos-Santos, I.M., \& al. (2016). The distribution of mass components in simulated disc galaxies. Mon. Not. Roy. Astron. Soc. 455, 476–483.
\bibitem{scarpa06}Scarpa, R. (2006). Modified Newtonian Dynamics, an Introductory Review. AIP Conference Proceedings 822, 102, arXiv:astro-ph/0601478.
\bibitem{serra10}Serra, A.L., Angus, G.W., \& Diaferio, A. (2010). Implications for dwarf spheroidal mass content from interloper removal.
    Astron. Astrophys. 524, A16.
\bibitem{skordis09}Skordis, C. (2009). TOPICAL REVIEW: The tensor-vector-scalar theory and its cosmology. Class.  Quant. Grav. 26, 143001.
\bibitem{sz19}Skordis, C. \& Z{\l}o\'{s}nik, T. (2019). Gravitational alternatives to dark matter with tensor mode speed equaling the speed of light.
    Phys. Rev. D 100, 104013.
\bibitem{steinhrdt}Steinhardt, C.L., \& al. (2016). The Impossibly Early Galaxy Problem. Astrophys. J. 824, 21.
\bibitem{tenneti18}Tenneti, A. \& al. (2018). The Radial Acceleration Relation in Disk Galaxies in the MassiveBlack-II Simulation.
    Mon. Not. Roy. Astron. Soc. 474, 3125-3132.
\bibitem{the88}The, L.S. \& White, S.D.M. (1988). Modified Newtonian dynamics and the Coma cluster. Astron. J. 95, 1642-1646.
\bibitem{tian19}Tian, Y \& Ko, C.M. (2019). Halo acceleration relation. Mon. Not. Roy. Astron. Soc. Lett. 488, L41-L46.
\bibitem{timlin18}Timlin, J.D., \& al. (2018). The Clustering of High-Redshift ($2.9 \leq z \leq 5.1$) Quasars in SDSS Stripe 82. Astrophys. J. 859, 20.
\bibitem{tiret08}Tiret, O. \& Combes, F. (2008). Evolution of spiral galaxies in modified gravity. II. Gas dynamics. Astron. Astrophys. 483, 719-726.
\bibitem{tiret09}Tiret, O. \& Combes, F. (2009). MOND and the dark baryons. Astron. Astrophys. 496, 659-668.
\bibitem{verlinde17}Verlinde, E.P. (2017). Emergent Gravity and the Dark Universe. SciPost Phys. 2, 016 (2017).
\bibitem{wang19}Wang, T., \& al. (2019). A dominant population of optically invisible massive galaxies in the early Universe. Nature 572, 211-214.
\bibitem{weinstein12}Weinstein, G. (2012). Einstein's 1912-1913 struggles with Gravitation Theory: Importance of Static Gravitational Fields Theory. arXiv:1202.2791.
\bibitem{weinstein19}Weinstein, G. (2019). Why did Einstein reject the November tensor in 1912-1913, only to come back to it in November 1915?  Stud. Hist. Philos. Mod. Phys. 62, 98-122. (arXiv:1904.02008)
\bibitem{wu15}Wu, X. \& Kroupa, P. (2015). Galactic rotation curves, the baryon-to-dark-halo-mass relation and space-time scale invariance. Mon. Not. Roy. Astron. Soc. 446, 330-344.
\bibitem{zhao08}Zhao, H.S. (2008). An Uneven Vacuum Energy Fluid as $\Lambda$, Dark Matter, MOND and Lens. Modern Phys. Lett. A, 23, 555-568.
\bibitem{zfs07}Zlosnik, T.G., Ferreira, P.G., \& Starkman, G.D. (2007). Modifying gravity with the aether: An alternative to dark matter. Phys. Rev. D,  75, 044017.
\bibitem{zs17}Z{\l}o\'{s}nik, T.G. \& Skordis, C. (2017). Cosmology of the Galileon extension of Bekenstein's theory of relativistic modified Newtonian dynamics. Phys. Rev. D 95, 124023.
\end{thebibliography}
\end{document}